\documentclass[letterpaper,twocolumn,10pt]{article}

\usepackage{usenix2019_v3}

\usepackage{tikz}
\usepackage{amsmath}
\usepackage{comment}
\usepackage{afterpage}
\usepackage[normalem]{ulem}
\usepackage{balance}
\usepackage[utf8]{inputenc}
\usepackage{tablefootnote}

\usepackage{pseudocode}
\usepackage{float}
\usepackage{xcolor}
\usepackage{ulem}
\usepackage{arydshln}
\usepackage{authblk}
\usepackage{soul}
\setstcolor{blue}

\definecolor{darkgreen}{rgb}{0.078,0.667,0.016}

\definecolor{codegreen}{rgb}{0,0.6,0}
\definecolor{codegray}{rgb}{0.5,0.5,0.5}
\definecolor{codeblue}{rgb}{0.039, 0.070, 0.760}
\definecolor{backcolour}{rgb}{1.0, 1.0, 1.0}

\usepackage{pifont}

\usepackage{booktabs}
\usepackage[ruled,linesnumbered]{algorithm2e}
\usepackage{amsthm}
\usepackage{xurl}
\usepackage{hyperref}
\usepackage{xspace}
\usepackage{enumitem}
\usepackage{subcaption}
\usepackage{multirow}
\usepackage{bbding}
\usepackage{tablefootnote}
\usepackage{ragged2e}
\justifying

\newcommand{\para}[1]{\noindent \textbf{#1 }}

\newcommand{\sys}{Cortex\xspace}

\newcommand{\cache}{LCFU\xspace}

\def\Snospace~{\S{}}

\usepackage{etoolbox}
\makeatletter
\patchcmd{\maketitle}
	{\@maketitle}
	{\@maketitle\vspace{-6.5em}}
	{}
	{}
\makeatother

\begin{document}
\title{\sys: Achieving Low-Latency, Cost-Efficient Remote Data Access For LLM via Semantic-Aware Knowledge Caching}

\makeatletter
\renewcommand\AB@affilsepx{, \protect\Affilfont}
\makeatother

\author[1]{Chaoyi Ruan}
\author[2]{Chao Bi}
\author[3]{Kaiwen Zheng}
\author[1]{Ziji Shi}
\author[4,1]{Xinyi Wan}
\author[1]{Jialin Li}

\affil[1]{National University of Singapore}
\affil[2]{USTC}
\affil[3]{University of Toronto}
\affil[4]{Sea AI Lab}

\maketitle

\begin{abstract}
\noindent
Large Language Model (LLM) agents tackle data-intensive tasks such as deep research and code generation.
However, their effectiveness depends on frequent interactions with knowledge sources across remote clouds or regions.
Such interactions can create non-trivial latency and cost bottlenecks.
Existing caching solutions focus on exact-match queries, limiting their effectiveness for semantic knowledge reuse.

To address this challenge, we introduce \sys, a novel cross-region knowledge caching architecture for LLM agents.
At its core are two abstractions: \textit{Semantic Element} (SE) and \textit{Semantic Retrieval Index} (Seri).
A semantic element captures the semantic embedding representation of an LLM query together with performance-aware metadata such as latency, cost, and staticity.
Seri then provides a two-stage retrieval pipeline: a vector similar index with semantic embedding for fast candidate selection and a lightweight LLM-powered \textit{semantic judge} for precise validation.
Atop these primitives, \sys builds a new cache interface that includes a new semantic-aware cache hit definition, a cost-efficient eviction policy, and proactive prefetching.
To reduce overhead, \sys co-locates the smaller LLM judge with the main LLM using adaptive scheduling and resource sharing.
Our evaluation demonstrates that \sys delivers substantial performance improvements without compromising correctness.
On representative search workloads, \sys achieves up to a 3.6$\times$ increase in throughput by maintaining cache hit rates of over 85\%\, while preserving accuracy virtually identical to non-cached baselines.
\sys also improves throughput for coding tasks by 20\%, showcasing its versatility across diverse agentic workloads.
\end{abstract}

\section{Introduction}

\noindent
LLM agents~\cite{hong2023metagpt,zheng2024large,fei2024multimodal} have emerged as powerful tools that autonomously execute tasks such as deep research assistance~\cite{jin2025search,song2025r1,li2025search} and sophisticated code generation~\cite{cursor-code,claude-code,ma2025sql}. They rely on multi-step reasoning loops that
strategically query external knowledge from either private knowledge bases via retrieval-augmented generation (RAG) \cite{lewis2020retrieval} or cloud search services through API calls \cite{gunjan2012search,ray2025survey}, and they then synthesize the retrieved information step by step to produce a final solution.
Yet, this fundamental dependence on external data retrieval introduces a critical bottleneck to agent performance: high latency and cost of external knowledge access.

Compared to non-agent LLM inference, which relies mainly on GPU computation to generate text, LLM agents frequently invoke external data retrieval services/tools such as web search APIs~\cite{googleSearch2025,serpapi2025} or RAG backend services~\cite{turbopuffer2025,chroma2025}.
Often, the data source and agent models are located in different data centers or regions and are connected by a wide-area network (WAN).
Agent deployment thus faces two major challenges: high monetary cost and cross-region latency due to remote tool calling. For applications~\cite{googleaimode-report,googleaimode-estimate} handling 5-10 million daily queries, costs are substantial.
Given the price of each Google Search API call at \$0.005~\cite{searchapiprice}, the application would incur a monthly costs of \$1.5-4.5 million. Furthermore, accessing data sources in remote regions can lead to 300-500~ms end-to-end latencies~\cite{cross-region-latency}. Such high latency can impact user experience and disrupt multi-step workflows.

Prior works have considered applying caching to reduce agent cost. However, existing caching strategies are designed to accelerate LLM inference itself, rather than mitigating the cost and latency of external data retrieval.
Semantic prompt caches~\cite{bang2023gptcache, google-cache,yan2025contextcache,schroeder2025adaptive} like GPT-Cache~\cite{bang2023gptcache} store LLM outputs and reuse responses via prompt similarity to skip LLM generation.
However, they compare prompts in the embedding space rather than at the knowledge or tool boundary, and they lack validation to ensure that reused content remains correct for the current context or time. Traditional data storage caches, such as key-value store~\cite{rocksdb2025}, database~\cite{mysql2025} or file system~\cite{ceph2025}, store KV objects indexed by \textit{exact keys}.
These caches lack the ability to evaluate semantic equivalence between non-matching queries. Transformer KV-caches~\cite{kwon2023efficient,gim2024prompt,liu2024cachegen} store token KV states to accelerate model decoding, but their scope is limited to inference computation.

To address this gap, we propose a new paradigm, \textit{semantic-aware remote knowledge caching}, to address latency and cost bottlenecks in cross-region data access for LLM agents. Unlike prior solutions that focus on optimizing LLM inference computation (e.g., GPTCache) or relying on exact-match queries, our approach leverages LLMs' language understanding to intelligently cache and retrieve external data.
Knowledge caching is built on two core abstractions. We first define a \textit{semantic element} (SE) that encapsulates the agent's query, tool interactions, and the retrieved response.
SE is augmented with performance-aware metadata such as latency, cost, and staticity. 
We then propose \textit{Semantic Retrieval Index (Seri)}, a two-stage retrieval engine that combines 1) an Approximate Nearest Neighbor (ANN) search for high-recall candidate selection and 2) a lightweight LLM-based \textit{Semantic Judge} for precise validation, ensuring true semantic equivalence.

We present \sys, a concrete implementation of semantic-aware knowledge caching. \sys seamlessly integrates the power of LLM semantic understanding into a robust caching architecture, significantly reducing external dependencies and costs associated with remote data access. Crucially, \sys bridges the gap between the uncertainty of semantic matching and the deterministic requirements of a traditional cache. This is achieved by transforming the ANN and semantic judge output into a reliable, semantic-aware cache hit signal. This procedure ensures that only genuinely relevant and contextually appropriate remote information is served. Furthermore, \sys equips the semantic index with advanced caching policies, such as a cost-efficient adaptive eviction policy and predictive prefetching, driven by SE metadata. Meanwhile, we propose an efficient \sys implementation using GPU co-location, where the primary agent LLM and the semantic judge LLM can reside on a single GPU managed by a priority-aware scheduler that protects the agent's critical latency paths.

Our evaluation demonstrates that \sys delivers substantial performance gains without compromising accuracy. On representative Zipfian and bursty workloads, \sys achieves up to a 3.6$\times$ throughput improvement over exact-match caching by sustaining cache hit rates of over 85\%. Crucially, our accuracy analysis reveals that. while a naive semantic cache suffers a significant drop in correctness, \sys maintains accuracy virtually identical to the non-cached baselines, proving the indispensable role of the semantic judge. This efficiency extends even to complex coding tasks, where \sys provides a 20\% increase in throughput.
\section{Background and Motivation}

\begin{figure*}[!t]
    \centering
    \begin{subfigure}[b]{0.32\textwidth}
        \centering
        \includegraphics[width=\textwidth]{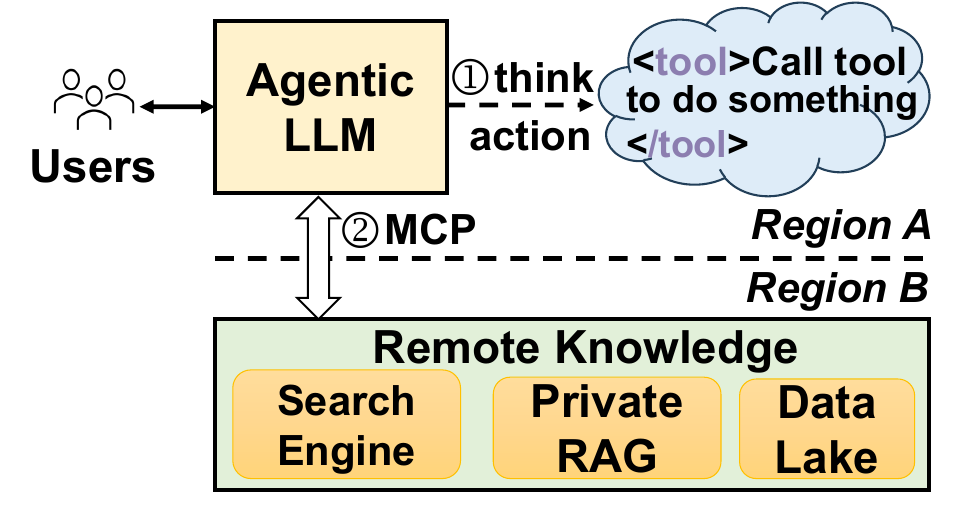}
        \caption{Agent-based application architecture.}
        \label{fig:agent-app}
    \end{subfigure}
    \hfill 
    \begin{subfigure}[b]{0.32\textwidth}
        \centering
        \includegraphics[width=1.01\textwidth]{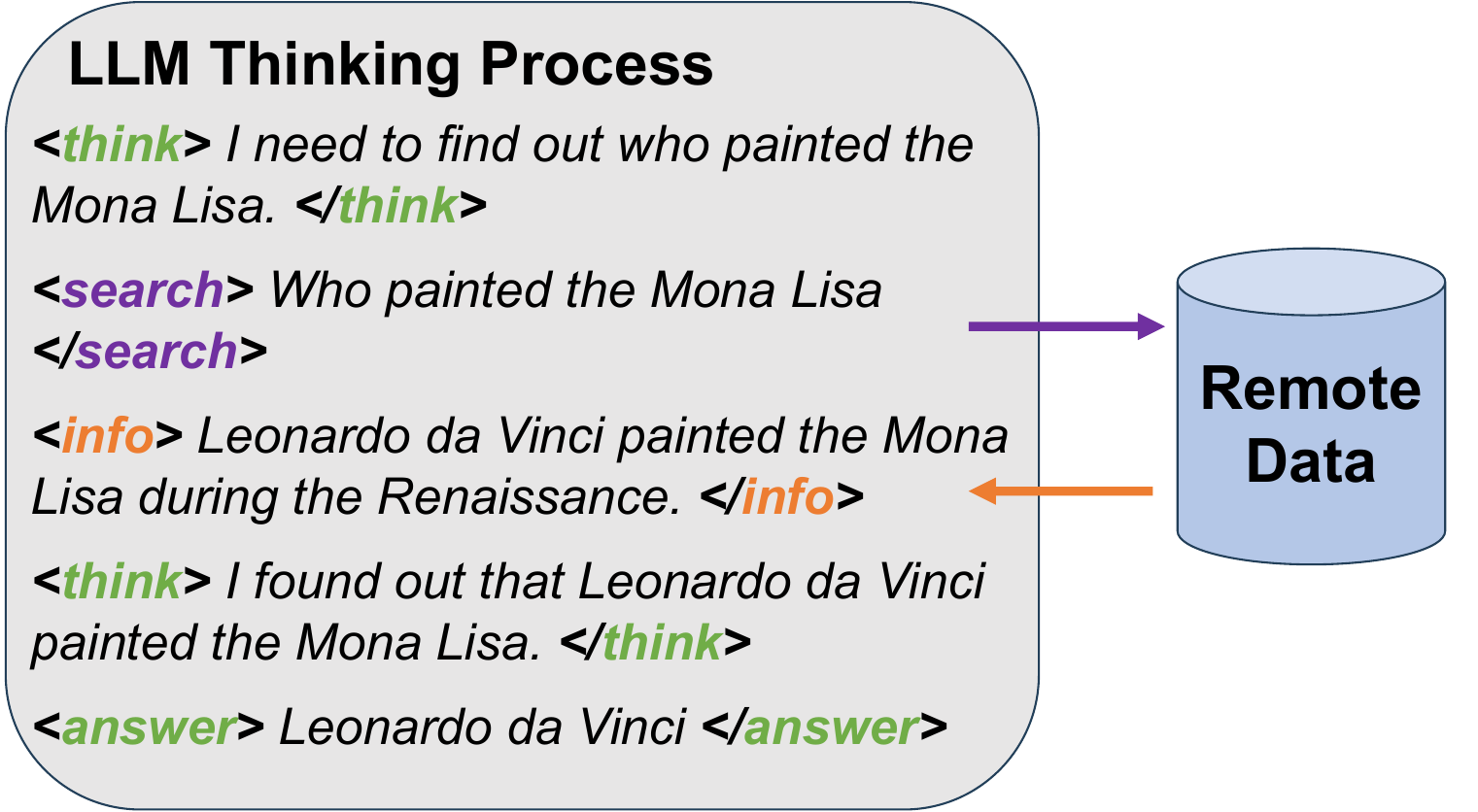}
        \caption{Example of Agentic LLM.}
        \label{fig:agent-case}
    \end{subfigure}
    \hfill 
    \begin{subfigure}[b]{0.33\textwidth}
        \centering
        \includegraphics[width=\textwidth]{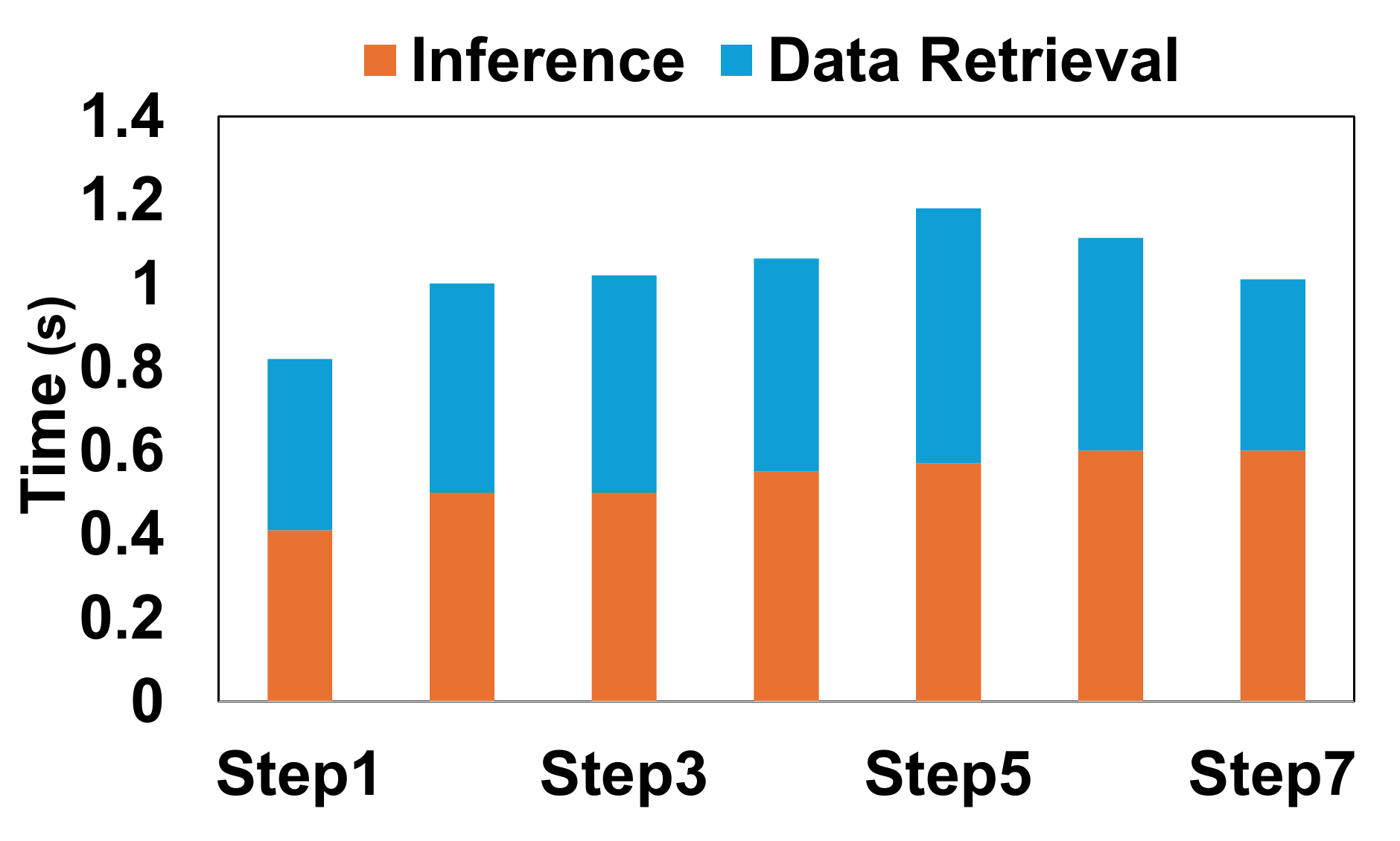}
        \caption{Search-R1 Latency breakdown.}
        \label{fig:agent-lat-break}
    \end{subfigure}
    \caption{Agentic workflow and analysis: (a) shows agent serving architecture which interacts with remote data service; (b) is a specific example from Search-R1~\cite{search_r1}; (c) presents breakdown of search queries when running Search-R1 7B on H100. }
\end{figure*}

\subsection{Agentic Applications Atop LLM}
\noindent
With the rise of reinforcement learning~\cite{deepseekai2025deepseekv3technicalreport,kimiteam2025kimik2openagentic}, LLMs are empowered with thinking capability, moving beyond simple text generation to generate tool-use commands/queries in intermediate reasoning steps. The most mature ones are searching agents~\cite{search_r1} and coding agents~\cite{claude-code}.

\para{Agentic LLM.} Unlike base LLM models~\cite{qwen25model}, primarily constrained by GPU inference speed, the core of an agentic system is an iterative ``think-act-observe'' loop that repeatedly accesses remote knowledge. As illustrated in~\autoref{fig:agent-app}, this process begins with a reasoning phase (\ding{172}, \textit{think}), where the agent formulates a plan. This plan then materializes when the LLM generates an \textit{action} (i.e., tool call query) to be dispatched over the model context protocol (MCP~\cite{antropic2024mcp}) to a remote data source. The tool call can be a search query or an API request. The dedicated \textit{think} and \textit{action} type of LLM outputs are encapsulated by special tags like \texttt{<think>} and \texttt{<tool>}. After retrieving remote knowledge, the agent integrates it into its working context to inform the next cycle. This loop of external data retrieval is not a one-shot step but a repeated operation at the heart of the agent's problem-solving process, distinguishing its operational profile from non-agentic LLMs. 

\para{The Search-R1~\cite{jin2025search} example.} To make this workflow concrete, consider the behavior of a search-focused agent as depicted in \autoref{fig:agent-case}. When tasked with finding the painter of the Mona Lisa, the agent first articulates its internal goal: \texttt{<think> }\textit{I need to find out who painted the Mona Lisa.} \texttt{</think>}. It then translates this thought into a concrete action, issuing a search tool call with the query \textit{who painted the Mona Lisa?}. This action triggers a remote tool call to an external search engine. Upon receiving the result, the agent observes it in an \texttt{<info>} block. Each \texttt{search} and \texttt{info} pair represents a single, costly round trip to an external data source.

\subsection{Performance and Cost Implications of Remote Data Retrieval.}
\label{subsec:tool-impl}

\begin{table}[!t]
\centering
\caption{Example cost information for the commonly used remote data access services.}
\resizebox{0.48\textwidth}{!}{
\begin{tabular}{lcccl}\hline
\toprule
\textbf{Company}& \multicolumn{1}{c}{\textbf{Google}} & \multicolumn{2}{c}{\textbf{OpenAI}}\\
\midrule
\textbf{Operation} & Search API  & Web Search Preview & Web Search\\
\textbf{Cost (per 1k reqs.)} & \$5~\cite{searchapiprice} & \$10-\$25~\cite{openaiapipricing} & \$10~\cite{openaiapipricing} \\
\bottomrule
\end{tabular}
}
\label{tab:remote_data_access}
\end{table}

\noindent
Heavy dependence on remote data sources creates significant latency and cost implications. Consider a standard cross-region deployment: an agent powered by a 7B model~\cite{jin2025search,jin2025model} runs on a single H100 GPU in one region while its external tools (e.g., a Search API) reside in another, introducing 100–300 ms network delay for every call. 

\para{Latency.} External retrieval often rivals the model's own inference time. As shown in \autoref{fig:agent-lat-break}, running Search-R1 on a single H100, external data retrieval constitutes around 40\%-50\% of the total execution time. While batching multiple requests can improve GPU utilization, the latency penalty of remote tool call is fundamental: each individual request must still wait for its remote tool calls to complete, as these calls lie on the critical path of the agent's reasoning loop.
This creates a latency floor that cannot be reduced via parallelism alone, directly impacting user-perceived response time.

\para{Financial cost.}
For large-scale deployments, API fees can become a substantial operational expense.
Take Google AI mode that handles 100 million MAU~\cite{googleaimode-report} as an example.
With an average of 10 tool calls per query~\cite{patil2023gorilla}, the service would incur \$0.005 per call, or roughly \$150,000 in daily API fees, as per \autoref{tab:remote_data_access}.
OpenAI charges up to \$25 per 1000 tool calls for Web Search functionality~\cite{openaiapipricing}, which is more expensive.
To serve such query volume with a production-grade dense model (e.g., Qwen2.5-72B or Llama-3-70B), a deployment requires approximately 50--100 H100 GPUs (at roughly \$1.49/hour each~\cite{hyperbolic2025}, or \$1,800--\$3,600/day). For this service, the \$150,000 daily API cost is 40--80$\times$ higher than that of GPU compute, making external data access the dominant operational expense.

\para{API rate limit.}
Beyond latency and cost, external API rate limits raise practical challenges for smaller-scale deployments.
For instance, commercial cloud APIs like Google Cloud Search~\cite{searchapiprice} impose strict rate limits, e.g., 100 queries per minute per user. Large enterprises can circumvent this by maintaining multiple accounts or negotiating higher quotas, but such solutions are impractical for individual developers and small teams.
As a result, many resort to third-party aggregation services (e.g., SerpAPI), which add both cost and external dependencies.
Cortex offers a lightweight alternative that reduces the number of external calls in the first place.

\subsection{Access Pattern of the Agentic Workloads}
\label{sec:access_pattern}
\noindent
Agentic LLM workloads heavily rely on remote tool calls, whose cost and latency often dominate performance. Caching can mitigate these overheads only if workloads exhibit sufficient locality. To assess this potential, we analyze two representative domains: AI-assisted search and code generation, both of which show statistical structures favorable to caching.

\para{Search-oriented agent workloads.}
Agentic search workloads are ultimately shaped by real-world human interests, which already drive global web search and are becoming more prominent as major engines adopt AI-powered modes~\cite{googleaimode2025,newbing2025}. Google Search, for instance, now incorporates agentic features—multi-source synthesis, conversational follow-ups, and contextual memory—that resemble the multi-step reasoning of LLM agents. Since commercial query logs are proprietary, we use public Google Trends data as a forward-looking proxy for the query distribution faced by large-scale agentic search systems.

\begin{figure}[!t]      \begin{center}\includegraphics[width=0.48\textwidth]{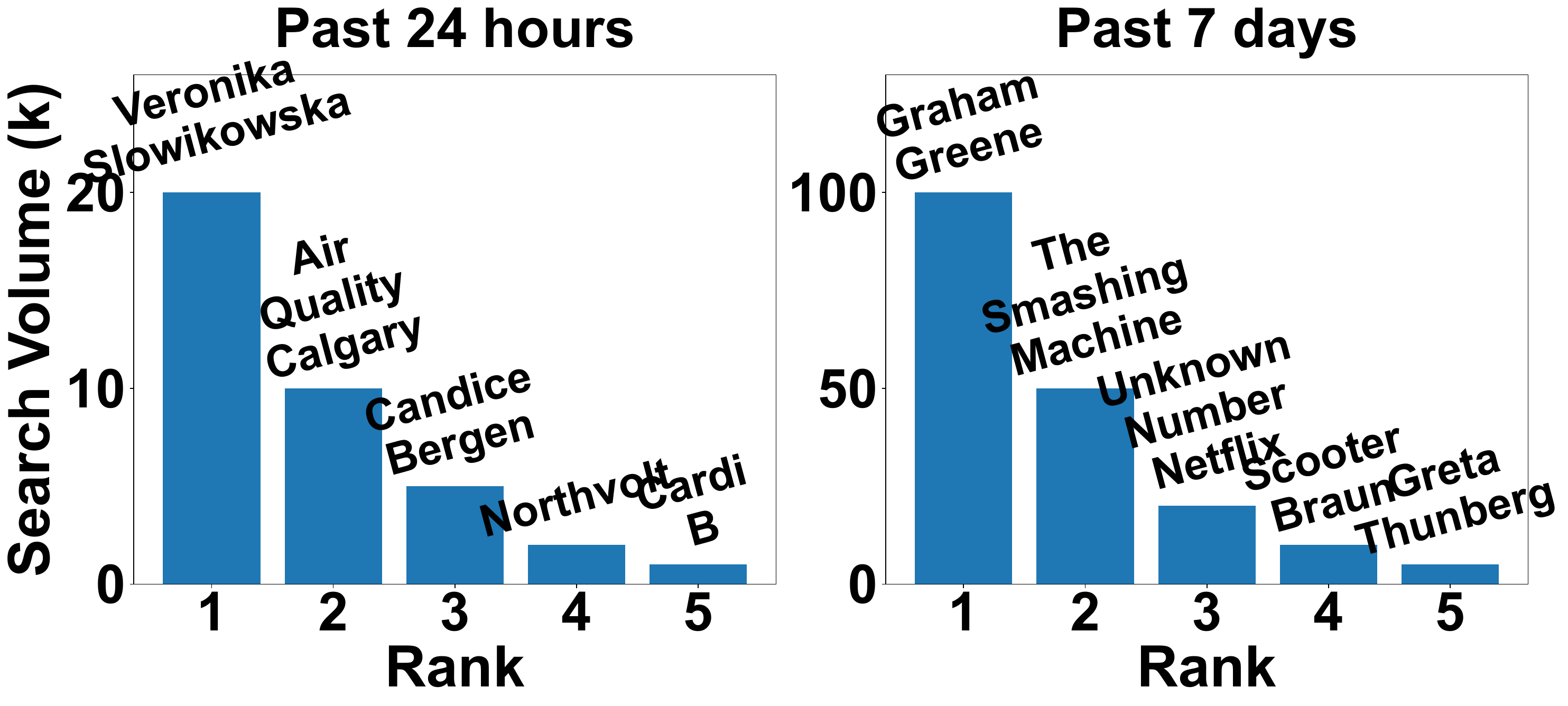}
    \caption{Google Trends data show that the top five topics over different time periods follow a Zipfian pattern. Approximate absolute volumes are used (e.g., ``20K+'' recorded as 20K), highlighting patterns in AI-assisted search and ranking.
    }
    \label{fig:google-trends-zipf}
    \end{center}
\end{figure}

\noindent
\textit{Zipfian Distribution of Search Interests.}
Search queries follow a Zipfian distribution: a few topics draw most traffic while the majority form a long tail. As shown in \autoref{fig:google-trends-zipf}, head queries like \textit{Veronika Slowikowska} and \textit{Graham Greene} dominate 24-hour and 7-day ranges, whereas thousands of others form the long tail. Caching these ``head'' topics can thus yield high hit rates with modest storage, greatly reducing remote calls.

\begin{figure}[!t]
    \begin{center}\includegraphics[width=0.48\textwidth]{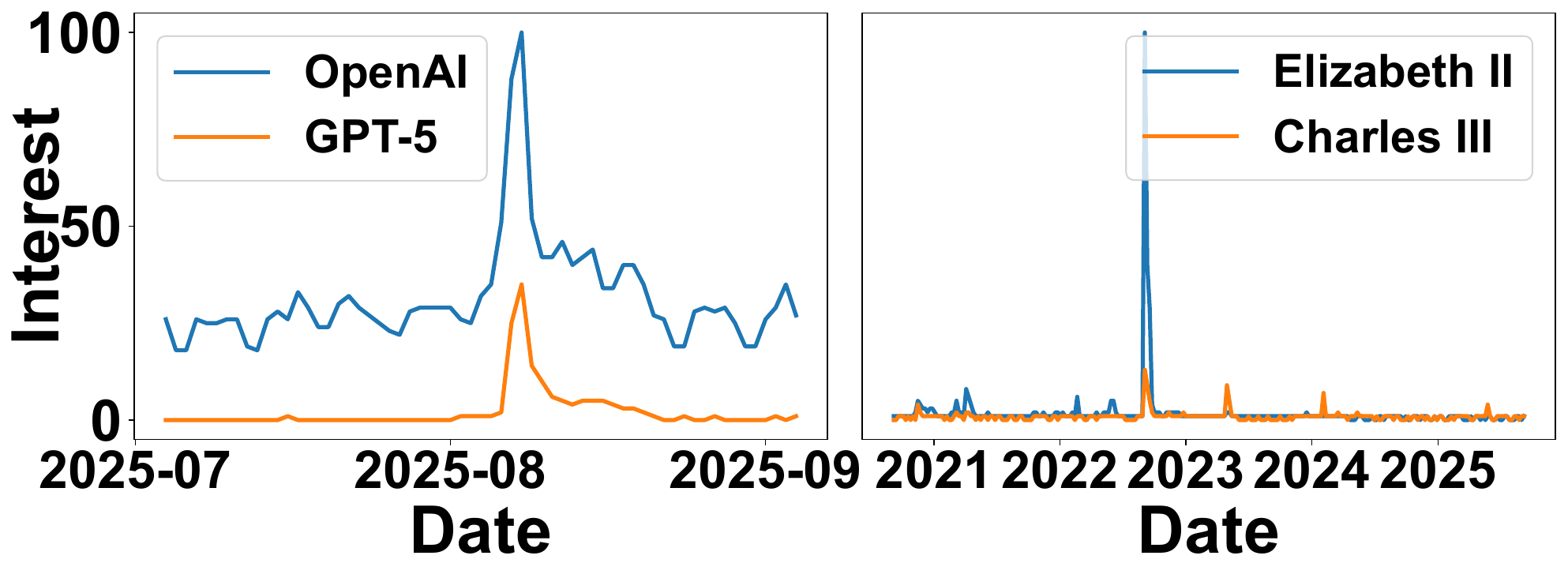}
    \caption{Empirical evidence of search queries resulting in bursty and correlated search patterns. A relevant external event leads to a surge in their search interest, as well as that of the topics with related themes.}
    \label{fig:google-trends-bursty}
    \end{center}
\end{figure}

\begin{table}[!t]
\centering
\resizebox{0.48\textwidth}{!}{
\begin{tabular}{lccccccccl}\hline
    \toprule
    \textbf{File-ID} & \textbf{1} & \textbf{2} & \textbf{3} & \textbf{4} & \textbf{5} & \textbf{6} & \textbf{7} & \textbf{8} & \textbf{9}\\
    \midrule
    \textbf{Access Freq.} & 1.0 & 0.28 & 0.22 & 0.14 & 0.1 & 0.08 & 0.04 & 0.04 & 0.04\\
    \bottomrule
\end{tabular}
}\caption{Access frequency of code file for SQLFluff~\cite{sqlfluff_repo} repo and coding problems are from SWE-Bench~\cite{swebench_oracle} Dev.}
\label{tab:code-agent-access}
\end{table}

\noindent
\textit{Bursty and Correlated Query Patterns.} Beyond a skewed global distribution, the temporal dynamics of queries reveal a second critical characteristic: they are often bursty and correlated. Interest in specific topics can experience sudden spikes in response to external events. As highlighted in \autoref{fig:google-trends-bursty}, these patterns range from technology breakthrough events (OpenAI releasing GPT-5) to major world news (the passing of Elizabeth II and Charles III's accession to the throne). The key implication for system design is twofold: caching strategies must be time-adaptive to absorb sudden spikes and can exploit topic correlations for proactive prefetching. 

\para{Code-agent workloads.}
While search-driven agents reflect global information demand, agentic LLMs are also widely used for code generation and software maintenance, which show a similarly skewed access pattern. To examine this, we analyzed coding tasks from SWE-Bench~\cite{swebench_oracle} on the \texttt{sqlfluff}~\cite{sqlfluff_repo} repository and measured how often each file is needed across tasks. As shown in \autoref{tab:code-agent-access}, file accesses follow a near-Zipfian distribution: one file is required by nearly all tasks, a few core modules are reused heavily, and most files are rarely touched. This long-tail pattern parallels the popularity skew seen in search workloads. This also implies that caching these hot files locally can eliminate many redundant cross-region fetches, and semantic matching can further capture requests that refer to the same file.

\subsection{Motivation and Cache Limitations} 
\label{subsec:moti}

\para{Motivation.}
Search-style and coding tasks show Zipfian popularity with bursty reuse, suggesting that caching could greatly reduce redundant remote calls. Yet, user queries in agentic workloads are semantically aligned rather than exactly key-matched, and they run in dynamic environments with complexities like cross-region latency, retrieval cost, and tool API limits.
A useful cache must handle \textit{semantic matching} among queries while accounting for \textit{cost}, \textit{cross-region latency}, and \textit{tool rate limits}.

\para{Existing caches and limitations.}
There are three common flavors of caching approaches, as summarized in \autoref{tab:cortex-comparison}, namely, transformer key-value (KV) caches~\cite{kwon2023efficient,gim2024prompt,liu2024cachegen}, semantic prompt caches~\cite{bang2023gptcache,google-cache,yan2025contextcache,schroeder2025adaptive}, and caches for traditional storage~\cite{rocksdb2025,mysql2025} or file system~\cite{ceph2025}. Despite their popularity, existing caching mechanisms exhibit significant drawbacks when handling agentic workloads.

\begin{table*}[!ht]
\centering
\caption{Cache comparison along critical requirements of agentic workloads.}
\resizebox{\textwidth}{!}{
\begin{tabular}{lccccc}
\toprule
\textbf{System} &
\textbf{Cached data} &
\textbf{Semantic Match} &
\textbf{Cost-aware Eviction} &
\textbf{Cross-region Aware} &
\textbf{Rate-limit aware} \\
\midrule
\textbf{Transformer KV-cache}~\cite{kwon2023efficient,gim2024prompt,liu2024cachegen} &
Token KV states &
\ding{55} & \ding{55} & \ding{55} & \ding{55} \\
\textbf{Semantic prompt cache}~\cite{bang2023gptcache,google-cache,yan2025contextcache,schroeder2025adaptive} &
LLM response &
\ding{55}
& \ding{55} & \ding{55} & \ding{55} \\
\textbf{Traditional storage cache}~\cite{rocksdb2025,mysql2025,ceph2025} &
KV object &
\ding{55} & \ding{51} & \ding{51} & \ding{55} \\
\textbf{\sys} &
External knowledge &
\ding{51} & \ding{51} & \ding{51} & \ding{51} \\
\bottomrule
\end{tabular}
}
\label{tab:cortex-comparison}
\end{table*}

The first type, transformer KV caches, accelerates token generation by reusing intermediate tensors for exactly-matched prompts, but its scope is confined to inference computation and does not extend to the remote data access layer.
The second type, semantic prompt caches (e.g., ContextCache~\cite{yan2025contextcache}, VectorQ~\cite{schroeder2025adaptive}), attempts to bypass inference entirely by matching user inputs via vector similarity. However, this approach suffers from a precision-recall trade-off: vector proximity often fails to capture true semantic equivalence. For example, VectorQ, despite applying adaptive thresholds for similar prompts, still relies on the limited semantic awareness of vector similarity, so a query like ``apple nutrition facts'' might incorrectly match ``Apple stock price analysis''. Moreover, these approaches are designed for plain LLM inference rather than agentic workloads. They cache model outputs at the prompt level but do not intercept or optimize external tool calls. For instance, ContextCache targets conversational LLMs and is not designed to handle the tool-calling patterns.
Lastly, studies have explored the data storage cache~\cite{park2024reducing} in databases or file systems in cross-region use cases. However, their exact-match mechanisms are fundamentally inadequate for natural language queries in agent workloads. These systems treat semantically equivalent queries as distinct keys, so even minor wording changes trigger cache misses and unnecessary remote calls.
Additionally, none of these caches fully addresses the rate limit impact in the agent scenario, so agent throughput can still be throttled even when cache hits.

\para{Takeaway.}
Agentic workloads require a cache that performs semantic match at the knowledge or tool boundary, validates correctness, is cost-aware, cross-region, and rate-limit aware so that cache hits yield real end-to-end gains. In contrast to these existing families, \sys satisfies all required properties as shown in \autoref{tab:cortex-comparison}, motivating its design in the next section.
\section{Design Overview}
\subsection{System Model and Goal}

\noindent
\para{Deployment model.} For LLM-based agents that rely on external knowledge and tools, frequent data retrieval from remote sources introduces significant latency and financial cost. Geographic separation between inference and data sources is often inevitable: organizations calling external LLM APIs cannot control where providers deploy their models, and even with local inference, agents frequently invoke external tools (e.g., web search, databases, third-party APIs) hosted in remote regions. In practice, agentic applications typically rely on the MCP protocol~\cite{antropic2024mcp} or other general RPC~\cite{grpc2025} over wide-area networks to fetch required data, amplifying these performance and cost penalties. To mitigate these overheads, we introduce \sys, a novel cross-region caching system designed to store and serve frequently accessed tool results from the perspective of agentic system efficiency.

\para{Optimization Goal.} The goal of \sys is to reduce costly remote data accesses by reusing semantically equivalent knowledge. Instead of relying on literal text similarity or exact matching, \sys leverages LLM-based semantic matching to identify and serve cached results for new queries with the same intent. This reduces query latency and significantly cuts API or tool usage costs, improving performance and operational efficiency in geo-distributed LLM agent deployments.

\subsection{Design Opportunities and Challenges}
\label{subsec:challenge}

\para{Strawman design.}
As discussed in \autoref{subsec:moti}, semantic prompt caches reuse model outputs based on embedding similarity. A natural strawman design is to extend this idea: embed each cached retrieval query, place all keys in an ANN index, and on a new query, return the value attached to its nearest neighbor. ANN methods~\cite{diskann,pq_ann} can search large embedding sets efficiently using graph structures or quantization. However, this naive design exposes key gaps. It treats similarity search as if it were a cache, yet it offers no guarantee that top-ranked items are actually correct, fresh, or cost-efficient to reuse. These issues motivate the following design challenges.

\para{Challenge 1: Ensuring accurate and valid semantic matches.}
Embeddings with ANN can reliably find and return textually similar queries, but textual or surface-level similarity~\cite{doostmohammadi2023surfacebasedretrievalreducesperplexity} does not imply semantic equivalence. Two queries can be near in the vector space yet require different answers or reflect different intents.
Correctness also depends on context and time. A match for a historical fact may remain valid, whereas a match for a fast-evolving topic can become stale. A practical system needs a validation mechanism that turns noisy similarity scores into trustworthy hit-or-miss decisions while accounting for both contextual fit and temporal validity.

\para{Challenge 2: From similarity search to a real cache.}
An embedding model and an ANN index are retrieval components, not a cache. They are typically used to search an entire corpus and return a ranked list of candidates. A cache, in contrast, is a capacity-limited, online component that must emit a deterministic hit signal, admit new items, evict old ones, respect freshness, and avoid polluting itself with near-misses. Using ANN as the front door of a cache forces several questions to be answered explicitly: when does a candidate become a cache hit rather than just a similar item; what metadata beyond the embedding are required to reason about staticity, cost, and scope; how should admission and eviction operate when items differ in lifetime, retrieval cost, and size; and how can we prefetch likely next items without undermining precision. Bridging this gap is essential to turn similarity search into a functional cache rather than a best-effort lookup.

\para{Challenge 3: Achieving sophistication with efficiency.}
Incorporating semantic validation, freshness reasoning, and predictive prefetching improves quality at the expense of higher compute and memory consumption. The system must integrate these capabilities while ensuring that the end-to-end savings from avoided remote calls outweigh local overheads, thus necessitating lightweight models, careful resource sharing with the serving LLM, and scheduling that protects user-facing latency.

\subsection{\sys Architecture Overview}
\begin{figure}[!t]
 \centering
 \includegraphics[width=0.44\textwidth]{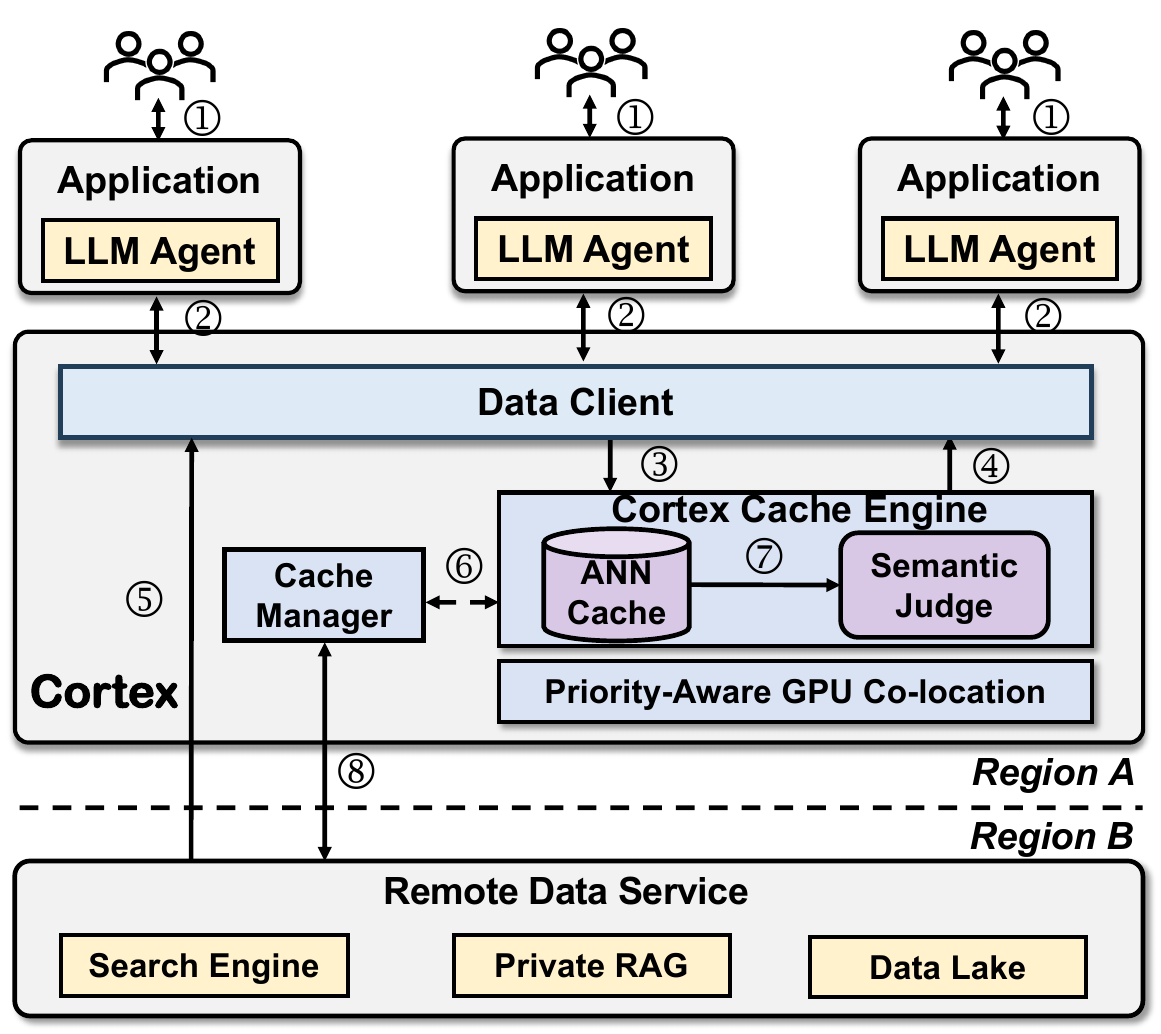}
 \caption{\sys's agentic caching architecture where agent LLM and data source reside in two geo-distributed regions }
 \label{fig:overview}
\end{figure}

\noindent
To address these three challenges, \sys adopts a layered architecture shown in \autoref{fig:overview}. 
Each component is explicitly designed to tackle a subset of these challenges while working together with other components as an integrated system. The top tier is the \textit{user agentic application}, where an LLM agent generates queries. The bottom tier is the Remote Data Service, the source of data, located in a remote region or cloud with a data retrieval latency around 300~ms-500~ms, following the common industrial practice~\cite{lin2024parrot}. This latency encompasses not only network RTT but also API processing and request queuing. Between these two lies the \sys Engine, which provides a transparent caching interface. It consists of three primary internal components:

\para{Data client.} The \textit{data client} serves as the transparent entry point into the \sys system, which intercepts all outgoing requests generated by the agent before they can reach external services. This interception allows \sys to seamlessly redirect the agent's queries into its internal caching workflow without modifying the agent application.

\para{\sys cache engine.}
At the core of the engine is the retrieval pipeline, engineered to achieve high-precision semantic matching. This pipeline operates in two distinct stages to balance speed with accuracy. Initially, an ANN index~\cite{diskann} (Faiss~\cite{douze2024faiss} is used in our experiments.) based cache rapidly identifies a set of candidate data items based on vector similarity. These candidates are then passed to a lightweight \textit{semantic judge} model (typically small with $\sim$1B parameters), which acts as a validation and is prompted to scrutinize each one for true contextual and semantic relevance to eliminate the false positives that plague naive semantic query caches.

\para{Agent-aware cache manager.}
Finally, the agent-aware cache manager provides the system's long-term intelligence by intelligently governing the cache's contents. Moving beyond simple heuristics, it employs two sophisticated strategies. Its adaptive eviction policy uses a utility-based model to decide which items to discard by considering factors like retrieval cost and importance to the agent's workflow. Complementing this, its proactive prefetching mechanism analyzes historical access patterns to proactively fetch data that is likely to be needed, ensuring the cache is dynamically optimized to hold the most valuable information.

\para{Workflow}
The workflow begins when the data client transparently intercepts a structured request and its surrounding context generated by the LLM agent (\ding{192}-\ding{193}). Instead of immediately dispatching the request externally, it first queries the \sys cache engine (\ding{194}). This query initiates a two-stage retrieval process designed for speed and accuracy. First, an ANN index rapidly surfaces a small set of candidate data items. Then, these candidates are scrutinized by the semantic judge (\ding{198}), ensuring accuracy. If the judge confirms a valid match, the cached result is immediately returned to the agent (\ding{195}). Otherwise, the request proceeds to the appropriate Remote Data Service (\ding{196}). The retrieved response is then returned to the agent and simultaneously stored in the cache as a new Semantic Element for future usage. Concurrently, the agent-aware cache manager observes the sequence of validated queries (\ding{197}), using this history to proactively prefetch data items (\ding{199}) and perform eviction, continuously optimizing the cache's contents in the background.
\section{\sys's Design}
\label{sec:mechanisms}

\noindent
This section details \sys's architecture for transforming agent interactions into reusable knowledge units with efficient caching, retrieval, and scheduling.

\subsection{Semantic Element}
\label{subsec:su}

\begin{figure}[!t]
    \centering
    \includegraphics[width=0.44\textwidth]{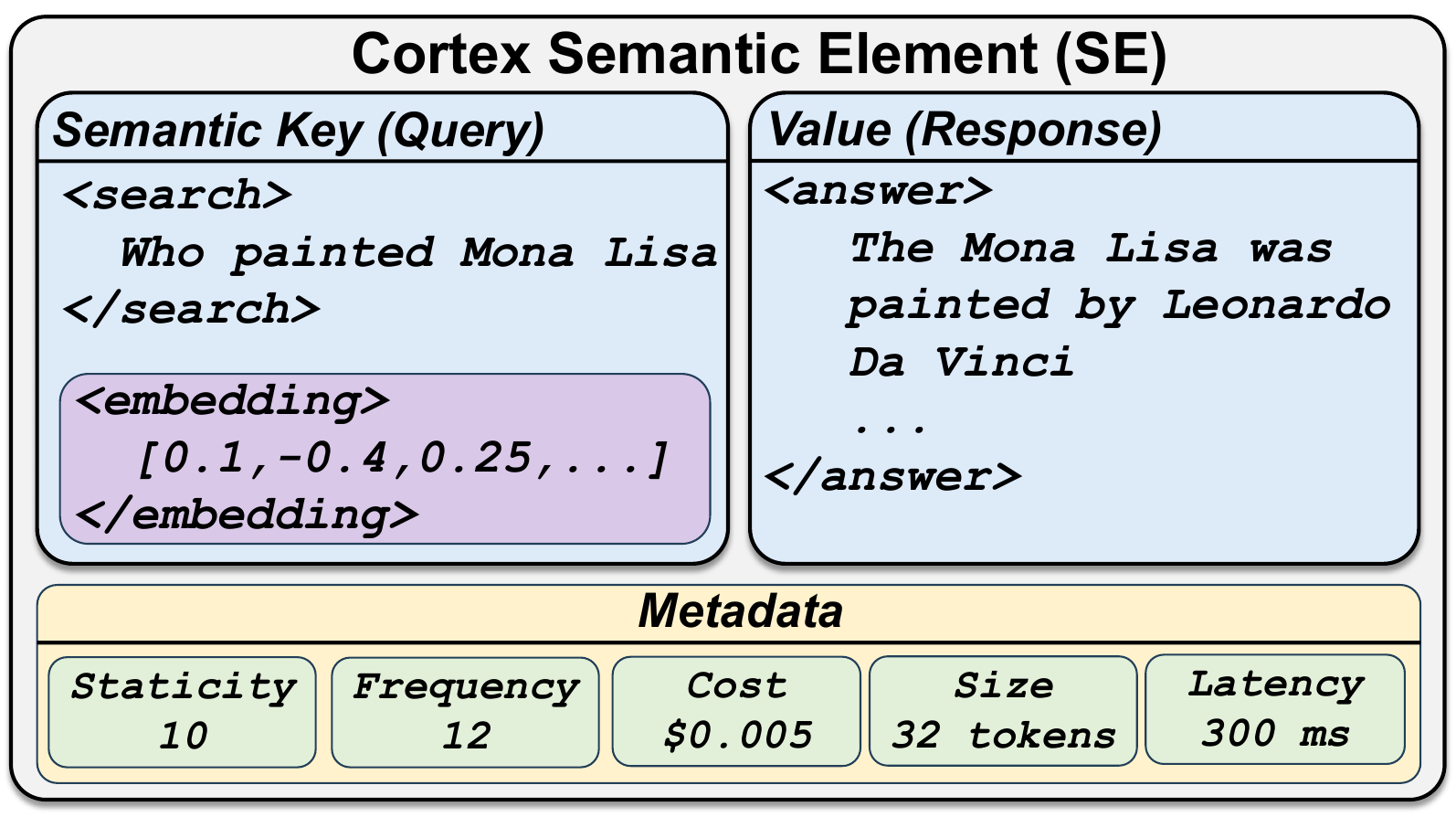}
    \caption{\sys's semantic element structure}
    \label{fig:semantic-element}
\end{figure}

\noindent
We begin with the \textit{Semantic Element (SE)}, which is \sys's core caching unit. 
An SE organizes an agent's \textit{query} and the \textit{retrieval result} for matching, validation, and reuse. As shown in \autoref{fig:semantic-element}, an SE is a key–value pair: the agent's query or tool action is the semantic key, and the retrieved information is the value. This structure leverages well-formed agentic outputs that wrap steps with tool tags. 
For example, the query ``Who painted the Mona Lisa'' within a \texttt{<search>} tag becomes the key, while the snippet in the \texttt{<info>} tag is the value. By reliably parsing these tagged blocks, \sys encapsulates discrete interactions (web searches, API calls, database lookups) into coherent SEs for caching.

An SE also carries metadata essential for lifecycle decisions. First, an embedding model (e.g., Qwen3-0.6B~\cite{qwen3_embedding_06b}) converts the semantic key into a high-dimensional vector used as a semantic fingerprint and is used for future matching. 
Concurrently, the semantic judge model (\autoref{subsec:retrieval}) estimates a \textit{staticity} score that quantifies the expected validity duration of the query–result pair on a scale of 1–10: higher scores (9--10) indicate stable facts that are unlikely to change, while lower scores (1--3) indicate time-sensitive information.
For instance, ``Who painted the Mona Lisa?'' is \textit{stable} (10), ``Who is the current US President?'' is \textit{moderate} (5), whereas ``Today's weather in Paris'' is \textit{ephemeral} (1).
This score impacts the Time-To-Live assignment and the eviction priority in \autoref{subsec:management}.

\subsection{Seri: Semantic Retrieval Index}
\label{subsec:retrieval}

\noindent
\sys retrieves semantic matches via via the \textit{Semantic Retrieval Index (Seri)}. 
Building on SEs, \sys implements a high-throughput, dual-stage retrieval pipeline engineered to resolve the fundamental trade-off between retrieval speed and semantic precision. This system is a sophisticated interplay of multiple specialized models, and its core is a formal constrained optimization problem: to minimize end-to-end query latency while rigorously maintaining semantic accuracy.

To formalize this, we define the components and their associated latencies.
The primary model is the agentic LLM, a powerful reasoning model that orchestrates tool use; its inference time is denoted as $L_{\text{Agent}}$. The cache itself relies on two smaller, highly optimized models: a lightweight \textit{Embedding LLM} for generating query vectors and a lightweight \textit{semantic judge model} (LSM), a classifier that validates whether a cached result answers a new query and estimates its staticity. The combined latency of a full two-stage cache lookup/check is $L_{\text{CacheCheck}} = L_{\text{ANN}} + L_{\text{LSM}}$. On a cache miss, an external tool call adds $L_{\text{Tool}}$.

\para{Coarse-grained filter.} The retrieval process begins with a high-recall candidate selection stage, where the threshold $\tau_{\text{sim}}$ (e.g., 0.9 in our setup) directly controls the scope of the search.
When the Agent LLM generates a new query, $q$, the system consults an ANN index. A candidate $c_i$ is selected if its similarity satisfies $\text{CosineSimilarity}(q, q_{\text{cached},i}) \ge \tau_{\text{sim}}$, where $q_{\text{cached},i}$ denotes the query of $c_i$. A lower, more permissive $\tau_{\text{sim}}$ increases recall, ensuring more potentially relevant items are passed to the next stage, but at the cost of increasing the validation workload. A higher, stricter $\tau_{\text{sim}}$ reduces this workload but risks prematurely discarding a correct match, thus lowering the potential hit rate.

\para{Fine-grained validation.} Selected candidates proceed to a high-precision semantic validation stage, where the LSM's threshold, $\tau_{\text{lsm}}$, is the primary lever for controlling accuracy (e.g., 0.9 in our experiment).
The LSM evaluates whether the cached result $r_{\text{cached},i}$ is a sufficient answer for the new query $q$, producing a confidence score $S_{\text{lsm}}$.
This guards against false positives from surface-level similarity, e.g., ``Apple nutrition facts'' vs.\ ``Apple stock price'' share keywords but require different answers.
The LSM rejects such matches by reasoning about semantic equivalence.
A cache hit is confirmed only if $S_{\text{lsm}} \ge \tau_{\text{lsm}}$. A higher $\tau_{\text{lsm}}$ enforces a stricter standard for equivalence, which increases the cache's precision (fewer false positives) but can decrease the hit rate by rejecting marginally correct answers. A lower $\tau_{\text{lsm}}$ boosts the hit rate but at the direct risk of serving more incorrect results, thereby lowering precision.

\para{Optimization goal.}
The system minimizes the expected latency subject to a target precision that is verified \textit{periodically offline}.
A hit incurs $L_{\text{hit}} = L_{\text{Agent}} + L_{\text{CacheCheck}}$, while a miss incurs $L_{\text{miss}} = L_{\text{Agent}} + L_{\text{CacheCheck}} + L_{\text{Tool}}$:
\[
\underset{\tau_{\text{sim}}, \tau_{\text{lsm}}}{\text{minimize}} \quad E[L] = P_{\text{hit}} \cdot L_{\text{hit}} + (1 - P_{\text{hit}}) \cdot L_{\text{miss}}
\]

The formula reflects an accuracy-throughput trade-off: lower $\tau_{\text{lsm}}$ improves hit rate but risks incorrect results; higher $\tau_{\text{lsm}}$ preserves accuracy but increases remote call frequency.

\para{Recalibration}
A fixed $\tau_{\text{lsm}}$ is brittle under workload drift or changing correctness requirements. \sys therefore performs periodic offline recalibration to adjust the LSM decision boundary while keeping the agent's latency-critical path unaffected.
This process, detailed in \autoref{alg:lsm_tuning}, begins by creating a high-quality annotated dataset from a sample of recent queries (\autoref{alg:line:start_annot}-\ref{alg:line:end_annot}).
Ground truth is obtained by re-issuing sampled queries to the live tool and comparing tool responses against the cached results.
The objective is then to find a new threshold, $\tau'_{\text{lsm}}$, that aligns the cache's performance on this dataset with a pre-defined metric, $P_{\text{target}}$.
This target precision represents the desired quality standard (e.g., 0.99 for 99\% correctness).
To find the new threshold, the system calculates a precision curve for the current LSM (\autoref{alg:line:predict}-\ref{alg:line:curve}) and finds the value that satisfies $P_{\text{target}}$ (\autoref{alg:line:find_thresh}), which is then deployed to the live system. Recalibration is decoupled from the request-serving path and does not block agent requests.

Crucially, this recalibration is highly cost-efficient: in our deployment, it samples 5 recent queries per minute, adding negligible overhead. The annotated set can also fine-tune the LSM to better capture subtle semantic differences over time.

\normalem
\begin{algorithm}[!t]
\caption{Periodic Threshold Recalibration}
\label{alg:lsm_tuning}
\KwData{Current LSM $J_{\text{lsm}}$, Target Precision $P_{\text{target}}$, Recent Eval Log $L_{\text{recent}}$, Validation Set $D_{\text{val}}$}
\KwResult{Recalibrated Threshold $\tau'_{\text{lsm}}$}

\DontPrintSemicolon
\SetAlgoVlined

$D_{sample} \leftarrow$ Sample diverse subset from $L_{\text{recent}}$\; \label{alg:line:start_annot}
$D_{\text{annotated}} \leftarrow \emptyset$\;
\For{$(q, r_{\text{cached}}, \_, \_) \in D_{\text{sample}}$}{
   $r_{\text{ground}} \leftarrow \text{FetchGT}(q)$\;
  $\text{label} \leftarrow \text{EvaluateGT}((q, r_{\text{cached}}), r_{\text{ground}})$\;
   $D_{\text{annotated}}$.Append($(q, r_{\text{cached}}, \text{label})$)\; \label{alg:line:end_annot}
}
$\text{scores} \leftarrow \text{PredictScores}(J_{\text{lsm}}, D_{\text{val}})$\; \label{alg:line:predict}
$\text{precision\_levels} \leftarrow \text{CalcPrecisionCurve}(\text{scores})$\; \label{alg:line:curve}
$\tau'_{\text{lsm}} \leftarrow \text{FindThreshold}(\text{precision\_levels}, P_{\text{target}})$\; \label{alg:line:find_thresh}
\text{UpdateSystem}($\tau'_{\text{lsm}})$\;
\KwRet{$\tau'_{\text{lsm}}$}\;
\end{algorithm}

\ULforem

\subsection{Building Cache Architecture Atop Seri}
\label{subsec:management}

\normalem

\begin{algorithm}[!t]
\caption{\cache Eviction Policy}
\label{alg:lcfu_eviction}
\KwData{Cache, capacity}
\DontPrintSemicolon
\SetAlgoVlined

\SetKwProg{Fn}{Function}{}{end}
\Fn{CalScore(se)}{
 ttl $\leftarrow$ se.ExpirationTime - CurrentTime()\;
 \If{se.Size == 0 or ttl <= 0}{
  \Return 0;
 }
\Return score $\leftarrow$ $\frac{\log(\text{se.Freq} + 1) \times \log(\text{se.Cost} \times 10^3 + 1) \times \log(\text{se.Lat} + 1) \times \log(\text{se.Stat} + 1)}{\text{se.Size}}$ \label{line:lcfu_score_calc}
}
Cache.RemoveExpired() \tcp{TTL purge first}
\If{Cache.Usage() > capacity}{
    \ForEach{se \textbf{in} Cache}{
        se.score $\leftarrow$ CalScore(se)\;
    }
    \While{Cache.Usage() > capacity}{
        victim $\leftarrow$ items.PopFirst()\; 
        Cache.Remove(victim)\;
    }
}
\end{algorithm}
\ULforem

\noindent
Although the Seri index can identify semantically matched items, it is not by itself a cache. To bridge the gap, \sys layers standard cache abstractions on top of the Seri, defining \textit{cache hit}, \textit{cache eviction} and \textit{cache prefetching} and translate semantic matching signals into cache behavior.

\para{Semantic-aware cache hit.} Unlike traditional caches where a hit is a simple key lookup, in our system a cached SE is only considered a ``hit'' after passing the full validation pipeline detailed in \autoref{subsec:retrieval}. On a new query, ANN retrieves a small set of candidates from the cache. Then the LSM (semantic judge) validates equivalence. Only a valid match registers as a cache hit, and that increments an SE's frequency count. This process effectively transforms a probabilistic similarity score into a definitive, binary hit/miss event, providing a stable, deterministic bedrock for a caching system.

\para{\cache eviction policy.}
With a reliable access signal established, we employ a tailored \textit{Least Cost-Efficient and Frequently Used} (\cache) eviction policy to manage the cache's contents. The insight behind \cache is that not all data are equally beneficial for cost and latency savings. Traditional LRU and LFU policies, which prioritize recency and frequency respectively, fail to account for the varying value of cached items. \cache addresses this by assigning each SE a \texttt{value\_score} that quantifies the benefit of retaining it.

As detailed in \autoref{alg:lcfu_eviction}, this \texttt{value\_score} is a composite metric derived from  retrieval  \texttt{latency} and \texttt{cost}, \texttt{frequency}, and \texttt{staticity}. Each factor captures a distinct retention benefit: \texttt{frequency} reflects reuse likelihood, \texttt{cost} and \texttt{latency} reflect savings per hit, and \texttt{staticity} reflects expected validity duration. Then the score is carefully normalized (\autoref{line:lcfu_score_calc}). 
It adjusts the raw metrics for frequency and cost to ensure they always contribute positively and meaningfully. This prevents new items or those with low costs from being unfairly penalized, given the fact that the cost per request is less than 1, and taking its logarithm will return a negative value. This combined score is then normalized by the item's size to provides a direct rationale: keep items that save the most time/money per byte. For instance, data with a high Retrieval Cost and high access Frequency will naturally receive a higher \texttt{value\_score}, making them less likely to be evicted. Conversely, ephemeral data with a lower Staticity will easily lead to lowering \texttt{value\_score}, making it a more likely candidate for eviction even if its frequency is high. For one Stable data, with a large \texttt{staticity} and high retrieval cost, it will maintain a higher \texttt{value\_score} even with less frequent access. This multi-attribute approach prevents transient but popular data from displacing enduring content, optimizing the cache for sustained performance and cost savings. When cache capacity is exceeded, items with the lowest \texttt{value\_score} are evicted, preserving the most valuable content. While a high \texttt{value\_score} indicates valuable content, such entries can become stale if kept indefinitely. To prevent outdated information from persisting, \sys integrates an aging mechanism using a user-defined TTL. Each cache entry is assigned a maximum lifespan, after which it is evicted regardless of its \texttt{value\_score}. This ensures that even high-cost or frequently accessed items are periodically refreshed, maintaining the cache's correctness while preserving the benefits of value-based retention.

\para{Predictive prefetching.} To complement the reactive \cache eviction policy, our framework employs proactive history-based prefetching to reduce miss latency. Using a lightweight first-order Markov model trained on confirmed cache hits, the system learns query-to-query transition patterns and calculates $P(q_{i+1}|q_i)$. When this probability exceeds a confidence threshold, the predicted query is asynchronously fetched and added to the cache.
Prefetched items enter with zero frequency, giving them minimal \texttt{value\_score}. If later requested and validated, their frequency increments and they become retained members. If unused, their low score makes them prime eviction candidates. This self-correcting design minimizes cache pollution from speculative fetches.

\subsection{Resource-Efficient Model Co-location}
\label{subsec:model-coloc}

\begin{figure}[!t]
    \centering
    \includegraphics[width=0.47\textwidth]{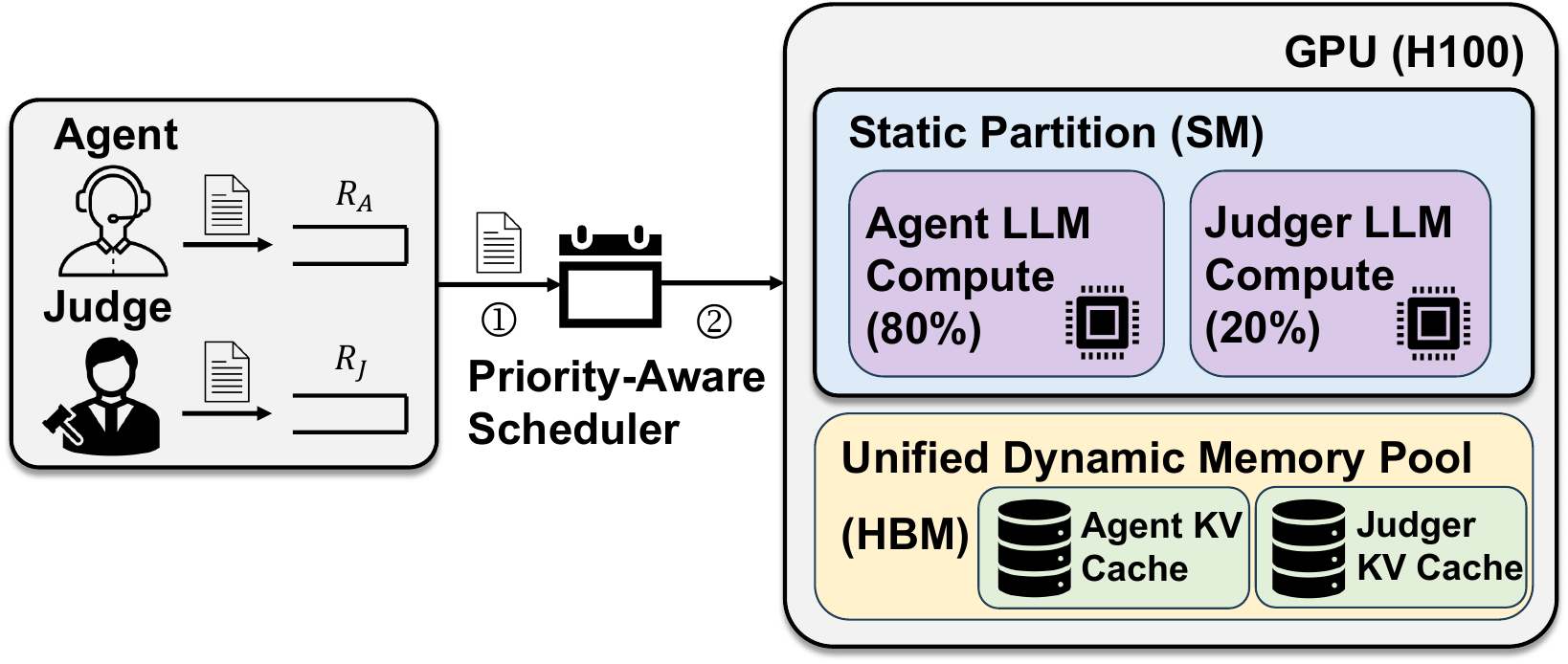}
    \caption{\sys's Priority-Aware Scheduling Procedure.}
    \label{fig:priority-aware-scheduling}
\end{figure}

\noindent
Although the two-stage retrieval computation is lightweight, the existing serving framework~\cite{vllm_repo} adopts one model per GPU mode. To keep overall latency low and achieve high-precision without doubling hardware costs, \sys co-locates the agent (e.g. $\sim$7B) and judge (e.g. $\sim$1B) efficiently on a single GPU. The challenge is protecting agent latency from judge processing.
Our co-location architecture (\autoref{fig:priority-aware-scheduling}) is built on a key insight: agent and judge workloads have fundamentally different priorities and resource profiles. Agent work is latency-critical, while judge work is a deferrable optimization. This is because a delayed validation does not block the user; at worst, it is treated as a cache miss, where performance for that single request degrades to the non-cached baseline of a full remote data call. 
Furthermore, the judge has a smaller memory demand since its prefill-only inference pattern (single token generation), results in a minimal and predictable KV cache footprint. Atop these principles, \sys builds a two-level defense, to ensure robust and efficient co-location.

\para{Coarse-grained asymmetric partitioning.} The foundation of our co-location strategy is to exploit the fundamentally asymmetric nature of the agent and judge workloads. The judge's workload profile is highly economical: unlike the agent which generates variable-length responses, the judge performs a classification task that yields a single token. This makes its primary consumer of GPU memory (i.e., the KV cache) minimal and highly predictable. Furthermore, its computational needs are lighter than the agent's complex, multi-step reasoning. To capitalize on this, we use the CUDA Multi-Process Service (MPS)~\cite{cuda-mps} to create a static, asymmetric compute partition, as shown in \autoref{fig:priority-aware-scheduling}. For instance, we allocate the dominant share of compute resources (e.g., 80\%) to the agent and a smaller, sufficient share (e.g., 20\%) to the judge. This coarse-grained partitioning optimizes for the common case, maximizing the performance of the primary agent LLM by dedicating the vast majority of resources to it.

\para{Fine-grained prioritization as a high-load guardrail.} While static partitioning is effective for the common case, a second, dynamic layer of defense is required to handle periods of high contention and guarantee the agent's low latency. This is the role of our fine-grained, priority-aware admission controller, which manages the unified dynamic memory pool ($M_{\text{dynamic}}$). This scheduler acts as a crucial guardrail by enforcing a strict prioritization policy. It services the agent queue ($Q_A$) exhaustively, only considering a batch from the judge queue ($Q_J$) when the agent queue is empty or lacks sufficient memory for the next dispatch. This ensures that the deferrable, internal work of the judge can never block the critical path of a user-facing agent task. This two-level defense combines asymmetric partitioning for baseline efficiency with dynamic priority scheduling for worst-case protection, enabling robust, high-performance co-location.
\section{Implementation and Discussion}

\noindent
\sys is implemented atop vLLM~\cite{vllm_repo}, intercepting tool calls and integrating with agent applications (e.g., Search-R1~\cite{search_r1}). For co-location, we use CUDA MPS to let multiple processes share a single GPU. Even when local tools are low-latency, \sys reduces cost by using semantic matching to avoid unnecessary tool calls and costs. We leave more advanced dynamic partitioning techniques (e.g., Green Contexts~\cite{cuda_green_contexts}) as future work.

\para{Judge accuracy.} Our framework bridges the gap between caching and semantic knowledge reuse by providing a unified cache abstraction for LLM agents, and it leverages powerful LLM-based techniques from prior research to ensure strong accuracy. The semantic judge can be easily fine-tuned or replaced for specific workloads, so its accuracy can be improved with minimal effort when needed. Given the rapid progress of small LLMs, we find this sufficient for practical use and view the judge as a pluggable component. In practice, \sys maintains high precision via strong semantic models and periodic recalibration, with misses falling back to live fetches to preserve correctness.

\section{Experiment}

\subsection{Experimental Setup}
\para{Testbed.}
We emulate a realistic cross-region deployment. The LLM agent and all \sys components run on an on-premise H100 cluster; remote data services run in a separate region. For search agent, we use the public Google Cloud Search API whose per-request average latency ranges between 300--500ms depending on response length, consistent with prior work~\cite{lin2024parrot}. For coding, we use a self-deployed FAISS~\cite{douze2024faiss}-based RAG service~\cite{douze2024faiss} with~300\,ms latency.

\para{Models and workloads.}
We evaluate two agentic scenarios: AI-powered search and code generation. The search agent is Search-R1-7B~\cite{jin2025search,jin2025model} (post-trained from Qwen-2.5 7B~\cite{qwen25model}); the coding agent is Qwen-3 8B~\cite{qwen3-8b}. \sys uses Qwen-3 0.6B models for embedding and semantic judging~\cite{qwen3_embedding_06b, qwen3_reranker_06b}. \sys's internal embedding and semantic judging functions are powered by the lightweight 0.6B models from the Qwen-3 family~\cite{qwen3_embedding_06b, qwen3_reranker_06b}.

\noindent
\textit{Search workloads.} Following the Google Trend access patterns in \autoref{sec:access_pattern}, we construct:
\textit{(i) Skewed workload}: extracting the Google-like head–tail access pattern in \autoref{sec:access_pattern}, we instantiate four search benchmarks--Zilliz-GPT~\cite{ziilz-gpt}, HotpotQA~\cite{hotpotqa_flashrag}, Musique~\cite{musique_flashrag}, and 2Wiki-Multi-Hop~\cite{2wikimultihopqa}. For each dataset, we apply $k$-means to the questions and keep 10 representative clusters, sampling $\sim$250 questions and constructing the skewed popularity per dataset (1000 total) and 
\textit{(ii) Trend-driven workload.} We capture 12-hour Google Trends time series for four topics, map them to HotpotQA questions, and compress them into a 10-minute trace to mimic sharp traffic spikes (\autoref{sec:access_pattern}).

\noindent
\textit{SWE-Bench workload.} We use a subset of widely-used SWE-Bench\_oracle~\cite{swebench_oracle} targeting the \texttt{sqlfluff} repo~\cite{sqlfluff_repo}, where the agent resolves GitHub issues. Each request represents one issue. This stresses caching because multiple issues repeatedly access shared files, such as core files and project documents.

\para{Baseline systems.}
To evaluate the end-to-end performance of an agent system, we compare three primary configurations, all built atop the vLLM framework~\cite{vllm_repo} for fair comparison. We compare:
\texttt{Agent\_vanilla} (no cache; every request hits the remote API),
\texttt{Agent\_exact} (traditional exact-match KV cache),
\texttt{Agent\_\sys} (our full system),
and an ablation \texttt{Agent\_ANN} that uses only ANN similarity. As \texttt{Agent\_ANN} is impractical for production (\autoref{subsec:challenge}), we restrict it to accuracy analysis (\autoref{subsec:accuracy}).

\para{Metrics.}
We report throughput (req/s), latency (ms), cache hit rate (\% served from cache), and operational cost (external API fees + GPU compute).

\begin{figure*}[!t]
    \centering
\includegraphics[width=0.9\textwidth]{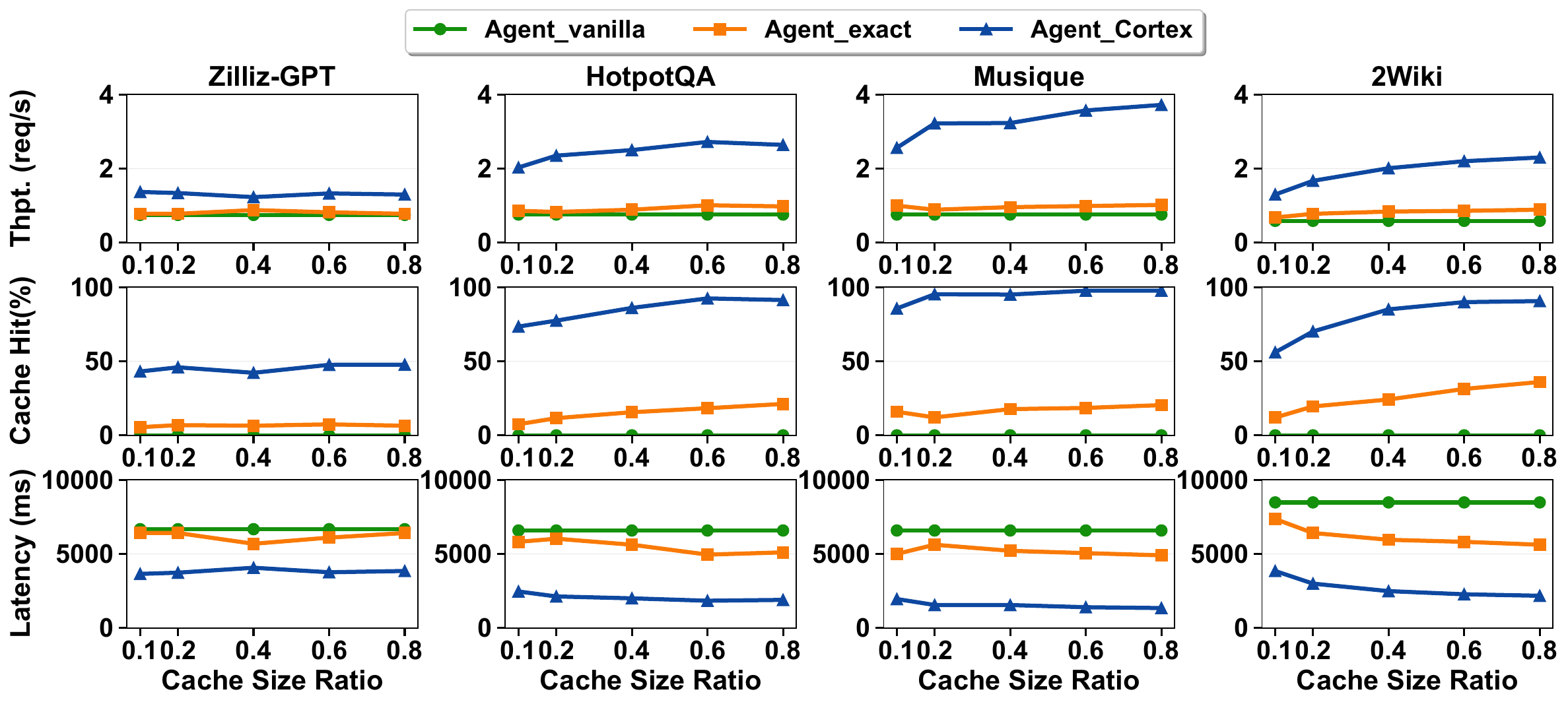}
    \caption{End-to-end agent serving throughput on skewed search workload under different cache ratio, zipfian-0.99 distribution}
    \label{fig:zipf_thput}
\end{figure*}

\subsection{Overall Performance On Cache Ratio}
\label{subsec:perf_cache}

\noindent
Using the setup above, we vary cache size ratio and report throughput, hit rate, and latency across three workloads.

\para{Skewed workload.}
We first evaluate \sys on search benchmarks with head–tail popularity skew (\autoref{sec:access_pattern}) to show semantic matching benefits. As shown in \autoref{fig:zipf_thput}, \sys outperforms the baselines across throughput, cache hit rate, and latency on all four datasets (Zilliz-GPT, HotpotQA, Musique, and 2Wiki). For instance, on Musique, Agent\_\sys achieves up to a 3.6$\times$ higher throughput: Agent\_\sys sustains over 85\% hit rates while the exact-match cache stays below 20\%. This high hit rate reduces end-to-end latency by up to 4$\times$ and minimizes reliance on external APIs, avoiding rate limits such as the Google Search API’s 100 queries per minute cap.
Note that absolute latencies exceed raw network RTTs because requests to external APIs experience queueing and backoff under rate limits, which inflates latencies, as will be shown in \autoref{subsec:perf_breakdown}.
This performance gap stems from semantic diversity. Agent\_exact fails on syntactic variations, yielding few cache hits, while Agent\_\sys groups semantically similar queries, converting semantic locality into high cache efficiency.

\begin{figure}[!t]
    \centering
\includegraphics[width=0.47\textwidth]{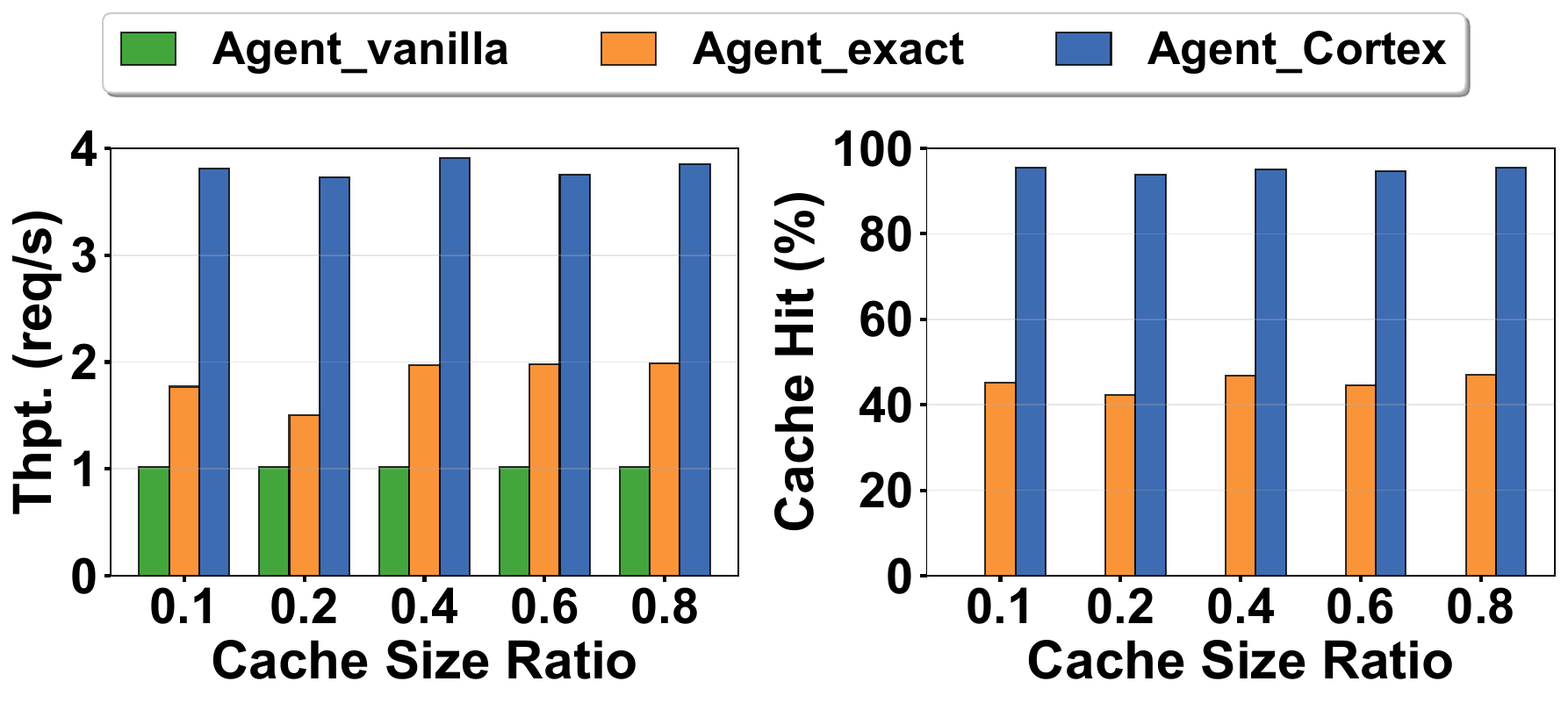}
    \caption{End-to-end agent serving throughput on trend-driven workload under different cache ratios}
    \label{fig:bursty_thput}
\end{figure}

\para{Trend-driven workload.}  We next evaluate Agent\_\sys on bursty workloads synthesized from Google Trends to test its adaptability. As shown in \autoref{fig:bursty_thput}, it achieves up to a 3.8$\times$ throughput improvement over the Agent\_vanilla baseline. This gain stems from its ability to maintain a high cache hit rate of nearly 95\%, effectively absorbing the intense but temporary demand of each trending topic.
This performance gain comes from the \cache eviction policy, which is tailored for temporal dynamics. Traditional caches retain obsolete data from past trends, reducing effective capacity. In contrast, \cache integrates an item's staticity into its priority score. As a volatile topic's popularity wanes, its cached entries are automatically deprioritized and become prime candidates for eviction. This mechanism ensures that cache space is continuously reclaimed for the next wave of content, which is crucial for maintaining high performance during bursty events.

\begin{figure}[!t]
    \centering
\includegraphics[width=0.47\textwidth]{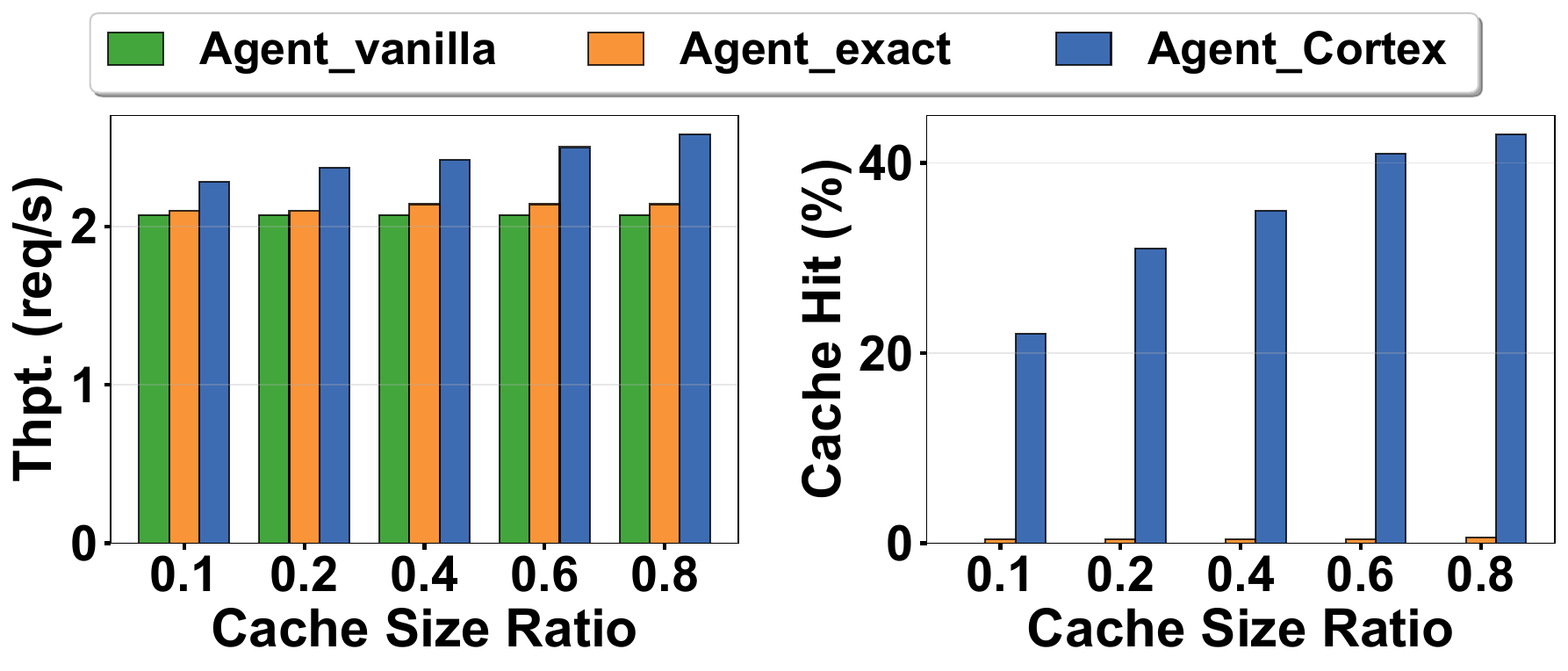}
    \caption{End-to-end agent serving throughput on SWE-Bench workload under different cache ratios}
    \label{fig:swe_thput}
\end{figure}

\para{SWE-Bench workload.} To demonstrate the generalizability of our approach, we evaluate  Agent\_\sys on a code generation workload using the SWE-Bench. As shown in \autoref{fig:swe_thput}, the results reveal a significant 20\% throughput improvement over both Agent\_vanilla and Agent\_exact, driven by a cache hit rate approaching 45\%. While more modest than the search results, these gains are highly impactful for the complex domain of software engineering and validate the versatility of our caching approach.
The caching opportunity in this domain arises from shared file dependencies across tasks, such as when an agent resolves multiple GitHub issues within the same repository. A traditional cache like Agent\_exact is ineffective because it treats requests for different parts of the same file as distinct misses. Agent\_\sys, however, semantically identifies the shared file context, enabling it to serve these requests from the cache. Although the inherent diversity of coding tasks leads to a lower overall hit rate compared to search, serving nearly half of all file access requests locally provides a substantial reduction in latency and system load.

\subsection{Scalability Under Concurrency}
\label{subsec:over_con}
\begin{figure}[!t]
    \centering
\includegraphics[width=0.45\textwidth]{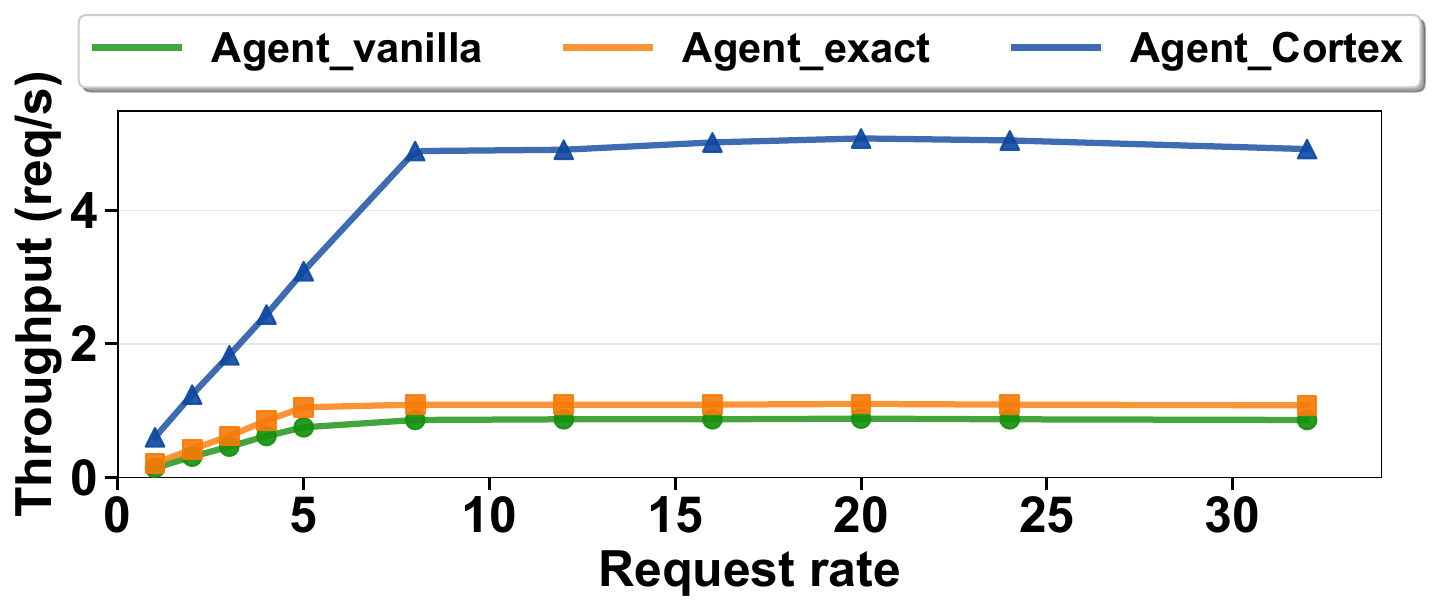}
    \caption{End-to-end Throughput Under Varying Request rate/concurrency on Musique Dataset}
    \label{fig:thput_con}
\end{figure}

\noindent
To assess the \sys's scalability under realistic workloads, we evaluate its end-to-end throughput with varying request concurrency. We use the Musique dataset at a fixed cache ratio of 0.4, a representative setting from our prior analysis. As shown in \autoref{fig:thput_con}, the results highlight \sys's significant scalability advantage, delivering up to a 5.7$\times$ and 4.5$\times$ throughput improvement comparing to the baselines. While the baseline systems quickly saturate, Agent\_\sys demonstrates strong scaling, achieving a throughput of 4.89 req/s at a request rate of 8. This represents a 4.5$\times$ improvement over Agent\_exact (1.09 req/s) and a 5.7$\times$ improvement over Agent\_vanilla (0.86 req/s) under heavy load. The throughput for both Agent\_vanilla and Agent\_exact plateaus early, at just around 1 req/s. This is because their performance is fundamentally bound by the remote data retrieval. With low cache hit rates, nearly every request involves a high-latency external API call, waiting for these remote responses. Increasing concurrency merely leads to longer queues, not higher throughput.
In contrast, Agent\_\sys's throughput scales nearly linearly with concurrency up to a request rate of 8, where high cache efficiency serves most requests locally and fully utilizes GPU parallelism—driving the agent to hardware-level capacity.
Beyond this point, \sys maintains stable throughput ($\sim$5 req/s) even as the request rate increases to 32, with only minor degradation ($<$3\%) at the highest load due to increased memory pressure.
The result demonstrates the robustness of our co-location design which provides sufficient resources for both the agent and the semantic judge without degrading to vanilla behavior under stress.

\subsection{Performance Breakdown}
\label{subsec:perf_breakdown}

\begin{figure}[t]
    \centering
    \includegraphics[width=0.45\textwidth]{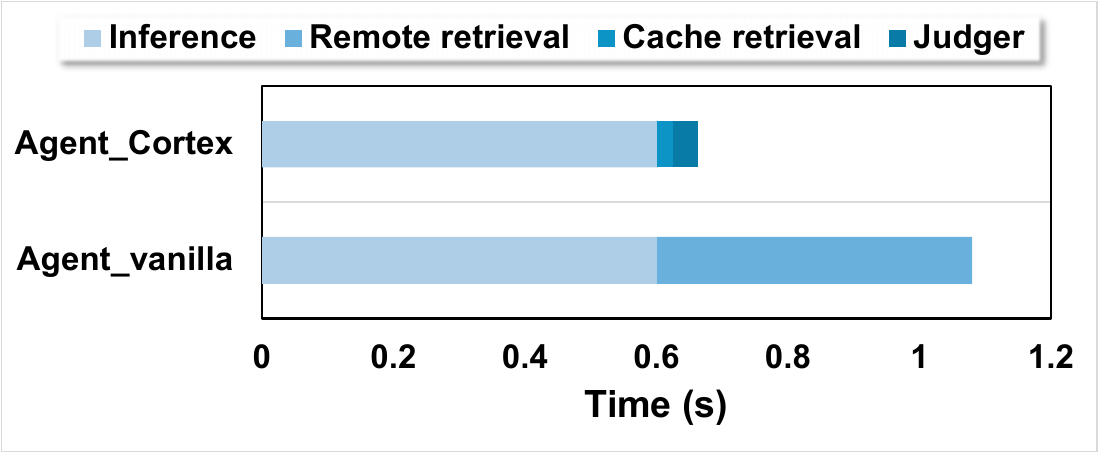}
    \caption{Per-request End-to-end Performance Breakdown}
    \label{fig:breakdown}
\end{figure}

\begin{figure}[t]
    \centering
    \includegraphics[width=0.44\textwidth]{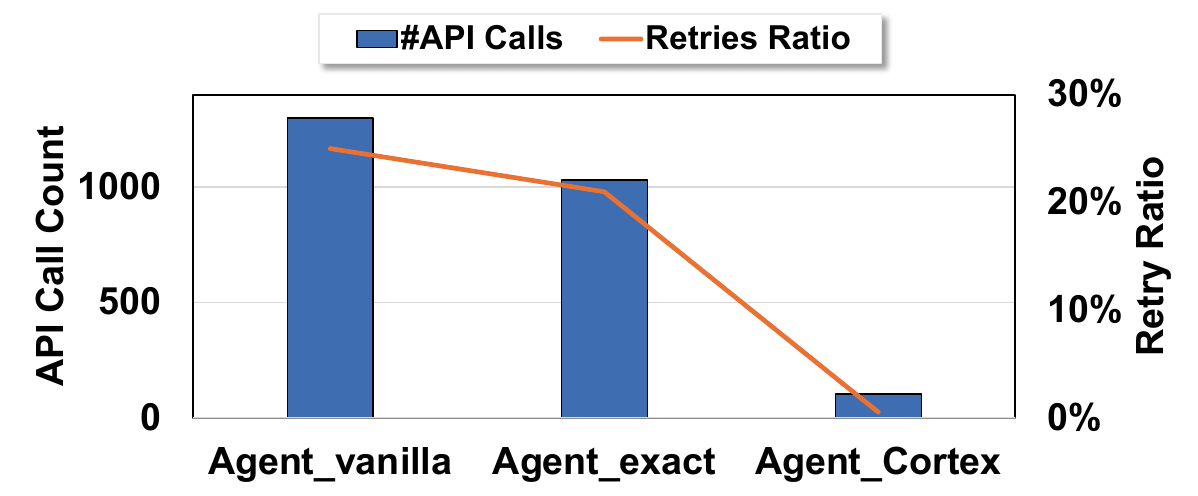}
    \caption{Data Retrieval Call and Retries Ratio}
    \label{fig:api_break}
\end{figure}

\para{Latency analysis.} 
To isolate pure request latency without rate-limit effects, we measure single-request breakdowns at low concurrency (\autoref{fig:breakdown}).
While agent inference remains 0.6~s, \sys cuts the total latency from 1.08~s to just 0.61s. This is achieved by almost entirely eliminating the 0.48~s external retrieval bottleneck that dominates the baseline's execution time. In its place, \sys introduces a minimal and predictable local overhead of only 0.05~s, comprising 0.02~s for cache retrieval and 0.03~s for judge validation. This trade-off validates our design principle: the marginal cost of intelligent local caching (0.05~s) is vastly outweighed by the savings from avoiding slow remote data fetches (0.48~s).

\begin{table}[!t]
\centering
\caption{Normalized throughput comparison between w/o. API Rate Limit and w. API Rate Limit}
\resizebox{0.45\textwidth}{!} {
    \begin{tabular}{ccc}\hline
    \toprule
    & \textbf{Without API Rate Limit} & \textbf{With API Rate Limit} \\
    \midrule
    \textbf{Agent\_vanilla} & 1& 1\\
    \textbf{Agent\_Cortex} & 1.5& 4.16\\
    \bottomrule
    \end{tabular}
    }
    \label{tab:limit_break}
\end{table}

\para{Cloud API rate limit and throughput analysis.} 
We then study throughput under realistic load where rate limits matter (\autoref{fig:api_break}). 
Beyond reducing single-request latency, \sys significantly enhances system throughput and scalability by mitigating the critical bottleneck of external API rate limits. This is achieved by fundamentally reducing the system's reliance on external services, as demonstrated in \autoref{fig:api_break}. Due to high cache miss rates, the non-cached Agent\_vanilla baseline generates approximately 1300 external API calls for the given task, leading to significant throttling and a high retry ratio of 25\%. In stark contrast, Agent\_\sys slashes the API call count to just 103, a 92\% reduction. This efficiency virtually cuts the retry ratio to a negligible 0.50\%.

To quantify how this API traffic reduction translates into a throughput advantage, we run a controlled experiment as shown in \autoref{tab:limit_break}. We use a self-deployed RAG data service with 300~ms here as we cannot cancel API rate limit for Google service. Even without rate limits, \sys already provides a 1.5$\times$ throughput improvement over the vanilla system due to its inherent latency savings. However, when a realistic API rate limit is enforced, the advantage becomes far more pronounced, with \sys achieving a 4.2$\times$ throughput gain over the throttled baseline. This comparison reveals that avoiding the rate-limiting bottleneck alone contributes an additional 2.8$\times$ improvement. By serving most requests locally, \sys sidesteps rate-limit bottlenecks and scales robustly.

\subsection{Cost Analysis}
\label{subsec:cost}

\begin{table}[!t]
\centering
\caption{Cost and performance comparison across different configurations.}
\resizebox{0.48\textwidth}{!}{
\begin{tabular}{lccc}
\toprule
\textbf{Metric} & \textbf{Agent\_vanilla} & \textbf{\sys w/o Sharing} & \textbf{\sys} \\ 
\midrule
\textbf{API Cost (\$)} & 6.5 & 6.5 & 0.64 \\ 
\textbf{GPU Cost (\$)} & 76 & 152 & 76 \\
\textbf{Total Cost (\$)} & 82.5 & 158.5 & 76.64 \\
\textbf{Thpt. (req/s)} & 0.87 & 4.74 & 4.89 \\
\textbf{Thpt./Cost (req/s/\$)} & 0.01 & 0.03 & 0.06 \\
\bottomrule
\end{tabular}
}
\label{tab:cost_performance_comparison}
\end{table}

\noindent
We now analyze cost-efficiency by evaluating three configurations under peak load on Musique (similar to \autoref{subsec:over_con}): Agent\_vanilla; \sys w/o Sharing, which uses an extra GPU for the semantic judge; and the complete \sys system. As detailed in \autoref{tab:cost_performance_comparison}, our analysis reveals that \sys is not only faster but fundamentally more economical, delivering 6$\times$ more throughput per dollar than the vanilla baseline.

This cost advantage stems from resolving the API–compute trade-off. The Agent\_vanilla system shows a total cost of \$82.5 to achieve a low throughput of 0.87 req/s. The \sys w/o Sharing configuration illustrates the pitfall of a naive caching approach. While dramatically boosting throughput to 4.79 req/s, it does so by requiring a dedicated second GPU for the semantic judge, doubling the GPU cost from \$76 to \$152 and making it the most expensive setup (\$158.5). In contrast, our co-located design cuts API cost by $>$90\% (to \$0.64) without extra hardware, keeping total spend at \$76.64 while retaining $\ge$95\% of the two-GPU throughput. Thus, \sys delivers $\sim$5.6$\times$ more performance at a similar cost, making it both fast and financially sustainable.

\begin{figure}[!t]
    \centering
\includegraphics[width=0.45\textwidth]{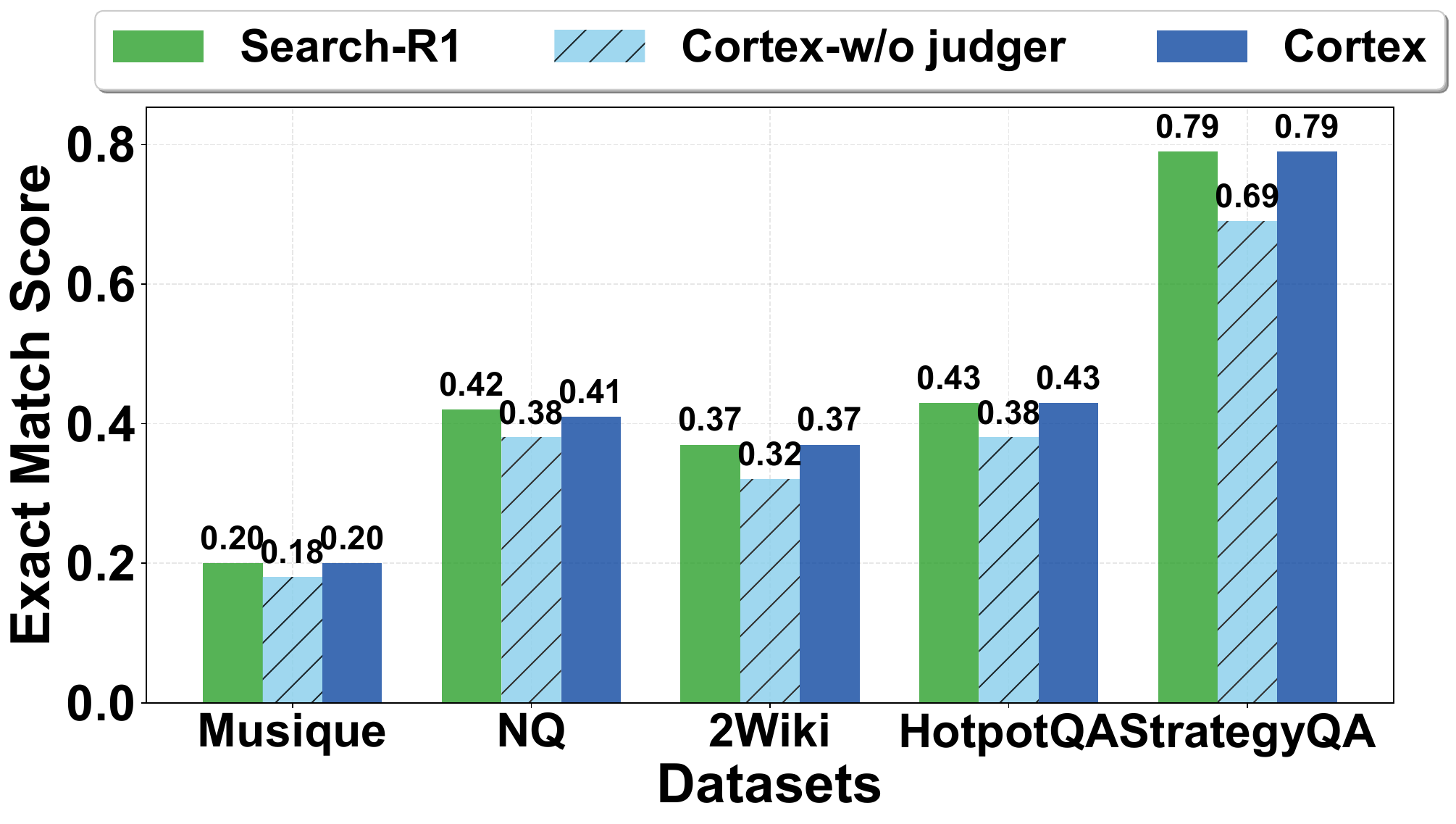}
    \caption{Generation quality comparison between \sys and vanilla Search-R1 (Agent\_vanilla)}
    \label{fig:acc-comparison}
\end{figure}

\subsection{Deep Dive}
\label{subsec:accuracy}

\para{Accuracy Analysis}
We measure correctness using Exact Match score~\cite{exactmatch2025,search_r1}. As shown in \autoref{fig:acc-comparison}, \sys-w/o judge shows accuracy drops across datasets (e.g., StrategyQA: 0.69 vs. 0.79), demonstrating that naive ANN similarity can return wrong results. The full \sys matches baseline accuracy—the Semantic Judge rejects non-equivalent candidates, confirming \sys achieves caching efficiency without sacrificing correctness.

\begin{table}[!t]
\centering
\footnotesize
\caption{The benefits brought by \cache.}
\begin{tabular}{lccc}
\toprule
\textbf{Metric} & \textbf{Agent\_LRU} & \textbf{Agent\_LFU} & \textbf{\cache} \\ 
\midrule
\textbf{Cache hit (\%)} & 0.88 & 0.89 & 0.86 \\ 
\textbf{Throughput (req/s)} & 2.14 & 2.16 & 2.35 \\
\bottomrule
\end{tabular}
\label{tab:lcfu_lat_cost}
\end{table}

\para{\cache study.}
We tested our \cache policy against LRU and LFU strategies on the HotpotQA workload, as shown in \autoref{tab:lcfu_lat_cost}. While LFU achieved a higher hit rate (0.89 vs our 0.86), our policy delivered higher performance with up to 9\% compared to LRU and LFU eviction policy, respectively. This trade-off is intentional: rather than maximizing hit rate, our policy prioritizes caching expensive-to-retrieve items, resulting in better system-wide latency reduction.

\begin{table}[!t]
\centering
\caption{Co-location efficiency on H100.}
\resizebox{0.45\textwidth}{!}{
\begin{tabular}{lcc}
\toprule
\textbf{Metric} & \textbf{Dedicated-2GPU} & \textbf{Co-located (MPS 80/20)} \\
\midrule
Throughput (req/s) & 2.89 & 2.72 \\
p99 latency (ms)   & 6601 & 7230 \\
\bottomrule
\end{tabular}
}
\label{tab:colocation_eff}
\end{table}

\para{Co-location study.} Co-locating the Agent and Semantic Judge on a single A100 via CUDA MPS (80/20), at a representative cache ratio (0.6), retains 94\% of dedicated throughput (2.72 vs.\ 2.89 req/s) with a 9.5\% increase in p99 latency, indicating a near-baseline efficiency.

\para{Recalibration overhead.}
We also quantify the cost of threshold recalibration for the semantic judge using HotpotQA/Musique. Compared to a variant without recalibration, throughput decreases by only 2\%. This bounded cost comes from brief offline validation on small samples to keep the precision target, stabilizing accuracy under drift and preserving high-precision hits. In practice, the overhead is negligible relative to \sys’s gains.
\section{Other Related Work}

\para{Semantic prompt caching.}
Recent LLM serving systems~\cite{bang2023gptcache,google-cache,yan2025contextcache,schroeder2025adaptive} like Google Apigee~\cite{google-cache} match via embeddings to bypass inference; ContextCache~\cite{yan2025contextcache} caches multi-turn responses; VectorQ~\cite{schroeder2025adaptive} adapts thresholds for similar prompts. 
However, these approaches are largely internal to the inference process and typically for single-cloud/node, leaving the cross-region remote data bottleneck unaddressed.
In contrast, \sys operates at the external-data layer, caching knowledge retrieved via tool calls to cut cross-region latency and API costs. The two approaches are thus orthogonal and can be deployed together for complementary benefits.

\para{Traditional and multi-cloud caching.}
Beyond semantic schemes, caches rely on LRU/LFU/TTL heuristics with Redis~\cite{redis-cache} or Memcached~\cite{memcached-cache}. 
Cross-region deployments introduce latency, consistency, and cost trade-offs. Examples include Macaron~\cite{park2024reducing} (auto-tuned tiers cutting data-lake costs under write-through), EVCache~\cite{netflix-cache} (eventual consistency via Kafka), Azure Cache for Redis (passive cross-region replication with client co-location), and Alluxio~\cite{li2018alluxio} (global namespace for on-demand hot data). These architectures assume exact-match keys, while \sys instead brings semantic-aware matching and adaptive policies to remote knowledge caching.

\para{LLM-based agent memory.}
Finally, agent memory systems~\cite{zhong2023memorybankenhancinglargelanguage,xu2025amemagenticmemoryllm,huang2023memorysandboxtransparentinteractive} improve LLM's reasoning by retaining past context. For example, MemoryBank~\cite{zhong2023memorybankenhancinglargelanguage} stores conversational summaries to track entities; A-MEM~\cite{xu2025amemagenticmemoryllm} generalizes the memory structure to diverse tasks; Memory Sandbox~\cite{huang2023memorysandboxtransparentinteractive} gives users explicit control. These systems give LLMs long-term memory to improve reasoning and choose the next action, but they don't reduce the cost of executing tool calls. \sys instead targets the external tool call itself, caching external results to make agent operation faster and cheaper.
\section{Conclusion}

\noindent
We present \sys, an agentic knowledge caching system designed to overcome the latency and cost challenges of LLM agents accessing external knowledge. \sys's novelty lies in its two-stage hybrid retrieval, combining ANN search with an LLM-powered Semantic Judger for semantic matching, and its resource-efficient co-location architecture. Our evaluation demonstrates that \sys significantly improves performance and maintains accuracy, offering a scalable and cost-effective solution for LLM agent deployments.

\section*{Acknowledgments}

\noindent
We thank our shepherd Shay Vargaftik and anonymous reviewers for their valuable comments.
This work is supported by the Singapore Ministry of Education, under Academic Research Fund Tier 1 grant T1 251RES2104.

\normalem

\raggedbottom
{\small
\bibliographystyle{plain}
\bibliography{ref}

@article{hong2023metagpt,
  title={Metagpt: Meta programming for multi-agent collaborative framework},
  author={Hong, Sirui and Zheng, Xiawu and Chen, Jonathan and Cheng, Yuheng and Wang, Jinlin and Zhang, Ceyao and Wang, Zili and Yau, Steven Ka Shing and Lin, Zijuan and Zhou, Liyang and others},
  journal={arXiv preprint arXiv:2308.00352},
  volume={3},
  number={4},
  pages={6},
  year={2023}
}

@article{zheng2024large,
  title={Large language models as reliable knowledge bases?},
  author={Zheng, Danna and Lapata, Mirella and Pan, Jeff Z},
  year={2024}
}

@inproceedings{fei2024multimodal,
  title={From multimodal llm to human-level ai: Modality, instruction, reasoning, efficiency and beyond},
  author={Fei, Hao and Yao, Yuan and Zhang, Zhuosheng and Liu, Fuxiao and Zhang, Ao and Chua, Tat-Seng},
  booktitle={Proceedings of the 2024 Joint International Conference on Computational Linguistics, Language Resources and Evaluation (LREC-COLING 2024): Tutorial Summaries},
  pages={1--8},
  year={2024}
}

@misc{cursor-code,
  author       = {{CURSOR}},
  title        = {The AI Code Editor},
  year         = {2025},
  howpublished = {\url{https://cursor.com/agents}},
  note         = {Accessed: 2025-07-22}
}

@misc{claude-code,
  author       = {{Anthropic}},
  title        = {Your code’s new collaborator},
  year         = {2025},
  howpublished = {\url{https://www.anthropic.com/claude-code}},
  note         = {Accessed: 2025-07-22}
}

@article{ma2025sql,
  title={Sql-r1: Training natural language to sql reasoning model by reinforcement learning},
  author={Ma, Peixian and Zhuang, Xialie and Xu, Chengjin and Jiang, Xuhui and Chen, Ran and Guo, Jian},
  journal={arXiv preprint arXiv:2504.08600},
  year={2025}
}

@article{jin2025search,
  title={Search-r1: Training llms to reason and leverage search engines with reinforcement learning},
  author={Jin, Bowen and Zeng, Hansi and Yue, Zhenrui and Yoon, Jinsung and Arik, Sercan and Wang, Dong and Zamani, Hamed and Han, Jiawei},
  journal={arXiv preprint arXiv:2503.09516},
  year={2025}
}

@article{song2025r1,
  title={R1-searcher: Incentivizing the search capability in llms via reinforcement learning},
  author={Song, Huatong and Jiang, Jinhao and Min, Yingqian and Chen, Jie and Chen, Zhipeng and Zhao, Wayne Xin and Fang, Lei and Wen, Ji-Rong},
  journal={arXiv preprint arXiv:2503.05592},
  year={2025}
}

@article{li2025search,
  title={Search-o1: Agentic search-enhanced large reasoning models},
  author={Li, Xiaoxi and Dong, Guanting and Jin, Jiajie and Zhang, Yuyao and Zhou, Yujia and Zhu, Yutao and Zhang, Peitian and Dou, Zhicheng},
  journal={arXiv preprint arXiv:2501.05366},
  year={2025}
}

@article{lewis2020retrieval,
  title={Retrieval-augmented generation for knowledge-intensive nlp tasks},
  author={Lewis, Patrick and Perez, Ethan and Piktus, Aleksandra and Petroni, Fabio and Karpukhin, Vladimir and Goyal, Naman and K{\"u}ttler, Heinrich and Lewis, Mike and Yih, Wen-tau and Rockt{\"a}schel, Tim and others},
  journal={Advances in neural information processing systems},
  volume={33},
  pages={9459--9474},
  year={2020}
}

@article{gunjan2012search,
  title={Search engine optimization with Google},
  author={Gunjan, Vinit Kumar and Kumari, Monika and Kumar, Amit and Rao, Allam Appa and others},
  journal={Asian Journal of Engineering and Applied Technology},
  volume={1},
  number={1},
  pages={36--42},
  year={2012}
}

@article{ray2025survey,
  title={A survey on model context protocol: Architecture, state-of-the-art, challenges and future directions},
  author={Ray, Partha Pratim},
  journal={Authorea Preprints},
  year={2025},
  publisher={Authorea}
}

@misc{google-cache,
  author       = {{Google Cloud}},
  title        = {Get started with semantic caching policies},
  year         = {2025},
  howpublished = {\url{https://cloud.google.com/apigee/docs/api-platform/tutorials/using-semantic-caching-policies#:~:text=policies%20to%20enable%20intelligent%20response,latency%2C%20and%20lower%20operational%20costs}},
  note         = {Accessed: 2025-07-22}
}

@inproceedings{bang2023gptcache,
  title={Gptcache: An open-source semantic cache for llm applications enabling faster answers and cost savings},
  author={Bang, Fu},
  booktitle={Proceedings of the 3rd Workshop for Natural Language Processing Open Source Software (NLP-OSS 2023)},
  pages={212--218},
  year={2023}
}

@inproceedings{park2024reducing,
  title={Reducing cross-cloud/region costs with the auto-configuring MACARON cache},
  author={Park, Hojin and Qiu, Ziyue and Ganger, Gregory R and Amvrosiadis, George},
  booktitle={Proceedings of the ACM SIGOPS 30th Symposium on Operating Systems Principles},
  pages={347--368},
  year={2024}
}

@misc{netflix-cache,
  author       = {{Netflix}},
  title        = {EVCache},
  year         = {2025},
  howpublished = {\url{https://github.com/Netflix/EVCache}},
  note         = {Accessed: 2025-07-22}
}

@book{li2018alluxio,
  title={Alluxio: A virtual distributed file system},
  author={Li, Haoyuan},
  year={2018},
  publisher={University of California, Berkeley}
}

@misc{redis-cache,
  author       = {{Redis}},
  title        = {Redis},
  year         = {2025},
  howpublished = {\url{https://redis.io/}},
  note         = {Accessed: 2025-07-22}
}

@misc{memcached-cache,
  author       = {{Memcached}},
  title        = {Memcached},
  year         = {2025},
  howpublished = {\url{https://memcached.org/}},
  note         = {Accessed: 2025-07-22}
}

@misc{cuda-mps,
  author       = {{NVIDIA}},
  title        = {Multi-Process Service},
  year         = {2025},
  howpublished = {\url{https://docs.nvidia.com/deploy/mps/index.html}},
  note         = {Accessed: 2025-07-22}
}

@misc{searchapiprice,
  author       = {{Google}},
  title        = {Custom Search API
},
  year         = {2025},
  howpublished = {\url{https://console.cloud.google.com/marketplace/product/google/customsearch.googleapis.com}},
  note         = {Accessed: 2025-08-22}
}

@misc{openaiapipricing,
  author       = {{OpenAI}},
  title        = {API Pricing},
  year         = {2025},
  howpublished = {\url{https://openai.com/api/pricing/}},
  note         = {Accessed: 2025-09-07}
}

@misc{cross-region-latency,
  author       = {{Sachin Agarwal}},
  title        = {Public Cloud Inter-region Network Latency as Heat-maps},
  year         = {2025},
  howpublished = {\url{https://medium.com/@sachinkagarwal/public-cloud-inter-region-network-latency-as-heat-maps-134e22a5ff19}},
  note         = {Accessed: 2025-07-22}
}

@inproceedings{kwon2023efficient,
  title={Efficient memory management for large language model serving with pagedattention},
  author={Kwon, Woosuk and Li, Zhuohan and Zhuang, Siyuan and Sheng, Ying and Zheng, Lianmin and Yu, Cody Hao and Gonzalez, Joseph and Zhang, Hao and Stoica, Ion},
  booktitle={Proceedings of the 29th symposium on operating systems principles},
  pages={611--626},
  year={2023}
}

@article{gim2024prompt,
  title={Prompt cache: Modular attention reuse for low-latency inference},
  author={Gim, In and Chen, Guojun and Lee, Seung-seob and Sarda, Nikhil and Khandelwal, Anurag and Zhong, Lin},
  journal={Proceedings of Machine Learning and Systems},
  volume={6},
  pages={325--338},
  year={2024}
}

@inproceedings{liu2024cachegen,
  title={Cachegen: Kv cache compression and streaming for fast large language model serving},
  author={Liu, Yuhan and Li, Hanchen and Cheng, Yihua and Ray, Siddhant and Huang, Yuyang and Zhang, Qizheng and Du, Kuntai and Yao, Jiayi and Lu, Shan and Ananthanarayanan, Ganesh and others},
  booktitle={Proceedings of the ACM SIGCOMM 2024 Conference},
  pages={38--56},
  year={2024}
}

@misc{xu2025amemagenticmemoryllm,
      title={A-MEM: Agentic Memory for LLM Agents}, 
      author={Wujiang Xu and Kai Mei and Hang Gao and Juntao Tan and Zujie Liang and Yongfeng Zhang},
      year={2025},
      eprint={2502.12110},
      archivePrefix={arXiv},
      primaryClass={cs.CL},
      url={https://arxiv.org/abs/2502.12110}, 
}

@misc{jin2025model,
  author       = {{Jin, Bowen and Zeng, Hansi and Yue, Zhenrui and Yoon, Jinsung and Arik, Sercan and Wang, Dong and Zamani, Hamed and Han, Jiawei}},
  title        = {SearchR1-nq\_hotpotqa\_train-qwen2.5-7b-em-ppo},
  year         = {2025},
  howpublished = {\url{https://huggingface.co/PeterJinGo/SearchR1-nq_hotpotqa_train-qwen2.5-7b-em-ppo}},
  note         = {Accessed: 2025-08-22}
}

@misc{qwen25model,
  author       = {{Alibaba Qwen Team}},
  title        = {Qwen/Qwen2.5-7B-Instruct},
  year         = {2025},
  howpublished = {\url{https://huggingface.co/Qwen/Qwen2.5-7B-Instruct}},
  note         = {Accessed: 2025-08-22}
}

@misc{qwen3-8b,
  author       = {{Alibaba Qwen Team}},
  title        = {Qwen/Qwen3-8B},
  year         = {2025},
  howpublished = {\url{https://huggingface.co/Qwen/Qwen3-8B}},
  note         = {Accessed: 2025-08-22}
}

@misc{ziilz-gpt,
  author       = {{zilliztech}},
  title        = {Zilliz GPT Cache Datasets},
  year         = {2025},
  howpublished ={\url{https://github.com/zilliztech/GPTCache/blob/main/examples/benchmark/similiar_qqp_full.json.gz}},
  note         = {Accessed: 2025-08-22}
}

@misc{swebench_oracle,
  author       = {Princeton NLP},
  title        = {SWE-bench Oracle Dataset},
  year         = {2023},
  howpublished = {\url{https://huggingface.co/datasets/princeton-nlp/SWE-bench_oracle}},
  note         = {Accessed: 2025-09-14}
}

@misc{zhong2023memorybankenhancinglargelanguage,
      title={MemoryBank: Enhancing Large Language Models with Long-Term Memory}, 
      author={Wanjun Zhong and Lianghong Guo and Qiqi Gao and He Ye and Yanlin Wang},
      year={2023},
      eprint={2305.10250},
      archivePrefix={arXiv},
      primaryClass={cs.CL},
      url={https://arxiv.org/abs/2305.10250}, 
}

@misc{huang2023memorysandboxtransparentinteractive,
      title={Memory Sandbox: Transparent and Interactive Memory Management for Conversational Agents}, 
      author={Ziheng Huang and Sebastian Gutierrez and Hemanth Kamana and Stephen MacNeil},
      year={2023},
      eprint={2308.01542},
      archivePrefix={arXiv},
      primaryClass={cs.HC},
      url={https://arxiv.org/abs/2308.01542}, 
}

@misc{antropic2024mcp,
      title={Introducing the Model Context Protocol}, 
      author={Anthropic},
      year={2025},
  howpublished = {\url{https://www.anthropic.com/news/model-context-protocol}},
  note         = {Accessed: 2025-08-22}
}

@misc{googleaimode-estimate,
    title={Google AI Mode’s Early Adoption and SEO Impact},
    author={Luke Harsel},
    howpublished={\url{https://www.semrush.com/blog/google-ai-mode-seo-impact/?utm_source=chatgpt.com}},
    note={Accessed: 2025-09-12}
}

@misc{googleaimode-report,    
    title={Google’s AI Overviews have 2B monthly users, AI Mode 100M in the US and India},
    author={Sarah Perez},
    howpublished={\url{https://techcrunch.com/2025/07/23/googles-ai-overviews-have-2b-monthly-users-ai-mode-100m-in-the-us-and-india/?utm_source=chatgpt.com}},
    note={Accessed: 2025-09-12}
}

@article{yan2025contextcache,
  title={ContextCache: Context-Aware Semantic Cache for Multi-Turn Queries in Large Language Models},
  author={Yan, Jianxin and Ni, Wangze and Chen, Lei and Lin, Xuemin and Cheng, Peng and Qin, Zhan and Ren, Kui},
  journal={arXiv preprint arXiv:2506.22791},
  year={2025}
}

@article{schroeder2025adaptive,
  title={Adaptive Semantic Prompt Caching with VectorQ},
  author={Schroeder, Luis Gaspar and Liu, Shu and Cuadron, Alejandro and Zhao, Mark and Krusche, Stephan and Kemper, Alfons and Zaharia, Matei and Gonzalez, Joseph E},
  journal={CoRR},
  year={2025}
}

@inproceedings{lin2024parrot,
  title={Parrot: Efficient serving of $\{$LLM-based$\}$ applications with semantic variable},
  author={Lin, Chaofan and Han, Zhenhua and Zhang, Chengruidong and Yang, Yuqing and Yang, Fan and Chen, Chen and Qiu, Lili},
  booktitle={18th USENIX Symposium on Operating Systems Design and Implementation (OSDI 24)},
  pages={929--945},
  year={2024}
}

@article{douze2024faiss,
      title={The Faiss library},
      author={Matthijs Douze and Alexandr Guzhva and Chengqi Deng and Jeff Johnson and Gergely Szilvasy and Pierre-Emmanuel Mazaré and Maria Lomeli and Lucas Hosseini and Hervé Jégou},
      year={2024},
      eprint={2401.08281},
      archivePrefix={arXiv},
      primaryClass={cs.LG}
}

@misc{search_r1,
  author       = {Bowen Jin and Hansi Zeng and Zhenrui Yue and Jinsung Yoon and Sercan Arik and Dong Wang and Hamed Zamani and Jiawei Han},
  title        = {Search-R1: An Efficient, Scalable RL Training Framework for Reasoning \& Search Engine Calling Interleaved LLMs},
  year         = {2025},
  howpublished = {\url{https://github.com/PeterGriffinJin/Search-R1}},
  note         = {Accessed: 2025-09-14}
}

@misc{vllm_repo,
  author       = {vLLM Project Contributors},
  title        = {vLLM: A High-Throughput and Memory-Efficient Inference and Serving Engine for LLMs},
  year         = {2025},
  howpublished = {\url{https://github.com/vllm-project/vllm}},
  note         = {Originally developed at Sky Computing Lab, UC Berkeley. Accessed: 2025-09-14}
}

@misc{qwen3_embedding_06b,
  author       = {Qwen Team},
  title        = {Qwen3-Embedding-0.6B: A Multilingual Text Embedding Model},
  year         = {2025},
  howpublished = {\url{https://huggingface.co/Qwen/Qwen3-Embedding-0.6B}},
  note         = {Supports over 100 languages and customizable embedding dimensions. Accessed: 2025-09-14}
}

@misc{qwen3_reranker_06b,
  author       = {Qwen Team},
  title        = {Qwen3-Reranker-0.6B: A Multilingual Text Reranking Model},
  year         = {2025},
  howpublished = {\url{https://huggingface.co/Qwen/Qwen3-Reranker-0.6B}},
  note         = {Supports over 100 languages and customizable instructions for retrieval tasks. Accessed: 2025-09-14}
}

@misc{2wikimultihopqa,
  author       = {NLPIR Lab at RUC},
  title        = {2wikimultihopqa: A Multi-hop Question Answering Dataset from Wikipedia},
  year         = {2025},
  howpublished = {\url{https://huggingface.co/datasets/RUC-NLPIR/FlashRAG_datasets/tree/main/2wikimultihopqa}},
  note         = {Part of FlashRAG datasets. Accessed: 2025-09-14}
}

@misc{hotpotqa_flashrag,
  author       = {NLPIR Lab at RUC},
  title        = {HotpotQA (FlashRAG Version): Multi-hop Question Answering Dataset},
  year         = {2025},
  howpublished = {\url{https://huggingface.co/datasets/RUC-NLPIR/FlashRAG_datasets/tree/main/hotpotqa}},
  note         = {Part of FlashRAG datasets. Accessed: 2025-09-14}
}

@misc{musique_flashrag,
  author       = {NLPIR Lab at RUC},
  title        = {Musique (FlashRAG Version): Multi-hop Question Answering Dataset},
  year         = {2025},
  howpublished = {\url{https://huggingface.co/datasets/RUC-NLPIR/FlashRAG_datasets/tree/main/musique}},
  note         = {Part of FlashRAG datasets. Accessed: 2025-09-14}
}

@misc{sqlfluff_repo,
  author       = {SQLFluff Contributors},
  title        = {SQLFluff: A Modular SQL Linter and Auto-Formatter},
  year         = {2025},
  howpublished = {\url{https://github.com/sqlfluff/sqlfluff}},
  note         = {Supports multiple SQL dialects and templated code including Jinja and dbt. Accessed: 2025-09-14}
}

@article{patil2023gorilla,
  title={Gorilla: Large Language Model Connected with Massive APIs},
  author={Shishir G. Patil and Tianjun Zhang and Xin Wang and Joseph E. Gonzalez},
  year={2023},
  journal={arXiv preprint arXiv:2305.15334},
}

@misc{hyperbolic2025,
  author       = {Hyperbolic AI},
  title        = {Hyperbolic GPU Marketplace: On-Demand NVIDIA GPU Rentals},
  year         = {2025},
  howpublished = {\url{https://app.hyperbolic.ai/}},
  note         = {Accessed September 2025},
}

@misc{rocksdb2025,
  author       = {Facebook Database Engineering Team},
  title        = {RocksDB: A Persistent Key-Value Store for Fast Storage},
  year         = {2025},
  howpublished = {\url{https://github.com/facebook/rocksdb}},
  note         = {Accessed September 2025},
}

@misc{mysql2025,
  author       = {Oracle Corporation},
  title        = {MySQL: The World's Most Popular Open Source Database},
  year         = {2025},
  howpublished = {\url{https://www.mysql.com/}},
  note         = {Accessed September 2025},
}

@misc{ceph2025,
  author       = {Ceph Foundation},
  title        = {Ceph: The Future of Storage},
  year         = {2025},
  howpublished = {\url{https://ceph.io/en/}},
  note         = {Accessed September 2025},
}

@misc{deepseekai2025deepseekv3technicalreport,
      title={DeepSeek-V3 Technical Report}, 
      author={DeepSeek-AI and Aixin Liu and Bei Feng and Bing Xue and Bingxuan Wang and Bochao Wu and Chengda Lu and Chenggang Zhao and Chengqi Deng and Chenyu Zhang and Chong Ruan and Damai Dai and Daya Guo and Dejian Yang and Deli Chen and Dongjie Ji and Erhang Li and Fangyun Lin and Fucong Dai and Fuli Luo and Guangbo Hao and Guanting Chen and Guowei Li and H. Zhang and Han Bao and Hanwei Xu and Haocheng Wang and Haowei Zhang and Honghui Ding and Huajian Xin and Huazuo Gao and Hui Li and Hui Qu and J. L. Cai and Jian Liang and Jianzhong Guo and Jiaqi Ni and Jiashi Li and Jiawei Wang and Jin Chen and Jingchang Chen and Jingyang Yuan and Junjie Qiu and Junlong Li and Junxiao Song and Kai Dong and Kai Hu and Kaige Gao and Kang Guan and Kexin Huang and Kuai Yu and Lean Wang and Lecong Zhang and Lei Xu and Leyi Xia and Liang Zhao and Litong Wang and Liyue Zhang and Meng Li and Miaojun Wang and Mingchuan Zhang and Minghua Zhang and Minghui Tang and Mingming Li and Ning Tian and Panpan Huang and Peiyi Wang and Peng Zhang and Qiancheng Wang and Qihao Zhu and Qinyu Chen and Qiushi Du and R. J. Chen and R. L. Jin and Ruiqi Ge and Ruisong Zhang and Ruizhe Pan and Runji Wang and Runxin Xu and Ruoyu Zhang and Ruyi Chen and S. S. Li and Shanghao Lu and Shangyan Zhou and Shanhuang Chen and Shaoqing Wu and Shengfeng Ye and Shengfeng Ye and Shirong Ma and Shiyu Wang and Shuang Zhou and Shuiping Yu and Shunfeng Zhou and Shuting Pan and T. Wang and Tao Yun and Tian Pei and Tianyu Sun and W. L. Xiao and Wangding Zeng and Wanjia Zhao and Wei An and Wen Liu and Wenfeng Liang and Wenjun Gao and Wenqin Yu and Wentao Zhang and X. Q. Li and Xiangyue Jin and Xianzu Wang and Xiao Bi and Xiaodong Liu and Xiaohan Wang and Xiaojin Shen and Xiaokang Chen and Xiaokang Zhang and Xiaosha Chen and Xiaotao Nie and Xiaowen Sun and Xiaoxiang Wang and Xin Cheng and Xin Liu and Xin Xie and Xingchao Liu and Xingkai Yu and Xinnan Song and Xinxia Shan and Xinyi Zhou and Xinyu Yang and Xinyuan Li and Xuecheng Su and Xuheng Lin and Y. K. Li and Y. Q. Wang and Y. X. Wei and Y. X. Zhu and Yang Zhang and Yanhong Xu and Yanhong Xu and Yanping Huang and Yao Li and Yao Zhao and Yaofeng Sun and Yaohui Li and Yaohui Wang and Yi Yu and Yi Zheng and Yichao Zhang and Yifan Shi and Yiliang Xiong and Ying He and Ying Tang and Yishi Piao and Yisong Wang and Yixuan Tan and Yiyang Ma and Yiyuan Liu and Yongqiang Guo and Yu Wu and Yuan Ou and Yuchen Zhu and Yuduan Wang and Yue Gong and Yuheng Zou and Yujia He and Yukun Zha and Yunfan Xiong and Yunxian Ma and Yuting Yan and Yuxiang Luo and Yuxiang You and Yuxuan Liu and Yuyang Zhou and Z. F. Wu and Z. Z. Ren and Zehui Ren and Zhangli Sha and Zhe Fu and Zhean Xu and Zhen Huang and Zhen Zhang and Zhenda Xie and Zhengyan Zhang and Zhewen Hao and Zhibin Gou and Zhicheng Ma and Zhigang Yan and Zhihong Shao and Zhipeng Xu and Zhiyu Wu and Zhongyu Zhang and Zhuoshu Li and Zihui Gu and Zijia Zhu and Zijun Liu and Zilin Li and Ziwei Xie and Ziyang Song and Ziyi Gao and Zizheng Pan},
      year={2025},
      eprint={2412.19437},
      archivePrefix={arXiv},
      primaryClass={cs.CL},
      url={https://arxiv.org/abs/2412.19437}, 
}

@misc{kimiteam2025kimik2openagentic,
      title={Kimi K2: Open Agentic Intelligence}, 
      author={Kimi Team et al.and Yifan Bai and Yiping Bao and Guanduo Chen and Jiahao Chen and Ningxin Chen and Ruijue Chen and Yanru Chen and Yuankun Chen and Yutian Chen and Zhuofu Chen and Jialei Cui and Hao Ding and Mengnan Dong and Angang Du and Chenzhuang Du and Dikang Du and Yulun Du and Yu Fan and Yichen Feng and Kelin Fu and Bofei Gao and Hongcheng Gao and Peizhong Gao and Tong Gao and Xinran Gu and Longyu Guan and Haiqing Guo and Jianhang Guo and Hao Hu and Xiaoru Hao and Tianhong He and Weiran He and Wenyang He and Chao Hong and Yangyang Hu and Zhenxing Hu and Weixiao Huang and Zhiqi Huang and Zihao Huang and Tao Jiang and Zhejun Jiang and Xinyi Jin and Yongsheng Kang and Guokun Lai and Cheng Li and Fang Li and Haoyang Li and Ming Li and Wentao Li and Yanhao Li and Yiwei Li and Zhaowei Li and Zheming Li and Hongzhan Lin and Xiaohan Lin and Zongyu Lin and Chengyin Liu and Chenyu Liu and Hongzhang Liu and Jingyuan Liu and Junqi Liu and Liang Liu and Shaowei Liu and T. Y. Liu and Tianwei Liu and Weizhou Liu and Yangyang Liu and Yibo Liu and Yiping Liu and Yue Liu and Zhengying Liu and Enzhe Lu and Lijun Lu and Shengling Ma and Xinyu Ma and Yingwei Ma and Shaoguang Mao and Jie Mei and Xin Men and Yibo Miao and Siyuan Pan and Yebo Peng and Ruoyu Qin and Bowen Qu and Zeyu Shang and Lidong Shi and Shengyuan Shi and Feifan Song and Jianlin Su and Zhengyuan Su and Xinjie Sun and Flood Sung and Heyi Tang and Jiawen Tao and Qifeng Teng and Chensi Wang and Dinglu Wang and Feng Wang and Haiming Wang and Jianzhou Wang and Jiaxing Wang and Jinhong Wang and Shengjie Wang and Shuyi Wang and Yao Wang and Yejie Wang and Yiqin Wang and Yuxin Wang and Yuzhi Wang and Zhaoji Wang and Zhengtao Wang and Zhexu Wang and Chu Wei and Qianqian Wei and Wenhao Wu and Xingzhe Wu and Yuxin Wu and Chenjun Xiao and Xiaotong Xie and Weimin Xiong and Boyu Xu and Jing Xu and Jinjing Xu and L. H. Xu and Lin Xu and Suting Xu and Weixin Xu and Xinran Xu and Yangchuan Xu and Ziyao Xu and Junjie Yan and Yuzi Yan and Xiaofei Yang and Ying Yang and Zhen Yang and Zhilin Yang and Zonghan Yang and Haotian Yao and Xingcheng Yao and Wenjie Ye and Zhuorui Ye and Bohong Yin and Longhui Yu and Enming Yuan and Hongbang Yuan and Mengjie Yuan and Haobing Zhan and Dehao Zhang and Hao Zhang and Wanlu Zhang and Xiaobin Zhang and Yangkun Zhang and Yizhi Zhang and Yongting Zhang and Yu Zhang and Yutao Zhang and Yutong Zhang and Zheng Zhang and Haotian Zhao and Yikai Zhao and Huabin Zheng and Shaojie Zheng and Jianren Zhou and Xinyu Zhou and Zaida Zhou and Zhen Zhu and Weiyu Zhuang and Xinxing Zu},
      year={2025},
      eprint={2507.20534},
      archivePrefix={arXiv},
      primaryClass={cs.LG},
      url={https://arxiv.org/abs/2507.20534}, 
}

@misc{googleaimode2025,
  author       = {Google LLC},
  title        = {Google AI Mode: A New Way to Search, Whatever’s on Your Mind},
  year         = {2025},
  howpublished = {\url{https://search.google/ways-to-search/ai-mode/}},
  note         = {Accessed September 2025},
}

@misc{newbing2025,
  author       = {Microsoft Corporation},
  title        = {The New Bing: AI-Powered Assistant for Your Search},
  year         = {2025},
  howpublished = {\url{https://www.microsoft.com/en-us/edge/features/the-new-bing?form=MA13FJ}},
  note         = {Accessed September 2025},
}

@misc{grpc2025,
  author       = {gRPC Authors},
  title        = {Introduction to gRPC},
  year         = {2025},
  howpublished = {\url{https://grpc.io/docs/what-is-grpc/introduction/}},
  note         = {Accessed September 2025},
}

@misc{doostmohammadi2023surfacebasedretrievalreducesperplexity,
      title={Surface-Based Retrieval Reduces Perplexity of Retrieval-Augmented Language Models}, 
      author={Ehsan Doostmohammadi and Tobias Norlund and Marco Kuhlmann and Richard Johansson},
      year={2023},
      eprint={2305.16243},
      archivePrefix={arXiv},
      primaryClass={cs.CL},
      url={https://arxiv.org/abs/2305.16243}, 
}

@inproceedings{diskann,
 author = {Jayaram Subramanya, Suhas and Devvrit, Fnu and Simhadri, Harsha Vardhan and Krishnawamy, Ravishankar and Kadekodi, Rohan},
 booktitle = {Advances in Neural Information Processing Systems},
 editor = {H. Wallach and H. Larochelle and A. Beygelzimer and F. d\textquotesingle Alch\'{e}-Buc and E. Fox and R. Garnett},
 pages = {},
 publisher = {Curran Associates, Inc.},
 title = {DiskANN: Fast Accurate Billion-point Nearest Neighbor Search on a Single Node},
 url = {https://proceedings.neurips.cc/paper_files/paper/2019/file/09853c7fb1d3f8ee67a61b6bf4a7f8e6-Paper.pdf},
 volume = {32},
 year = {2019}
}

@ARTICLE{pq_ann,
  author={Jégou, Herve and Douze, Matthijs and Schmid, Cordelia},
  journal={IEEE Transactions on Pattern Analysis and Machine Intelligence}, 
  title={Product Quantization for Nearest Neighbor Search}, 
  year={2011},
  volume={33},
  number={1},
  pages={117-128},
  doi={10.1109/TPAMI.2010.57}}

@misc{turbopuffer2025,
  author       = {Turbopuffer Inc.},
  title        = {Turbopuffer: Serverless Vector and Full-Text Search Built on Object Storage},
  year         = {2025},
  howpublished = {\url{https://turbopuffer.com/}},
  note         = {Accessed September 2025},
}

@misc{googleSearch2025,
  author       = {Google LLC},
  title        = {Google Programmable Search Engine: Custom Search Powered by Google},
  year         = {2025},
  howpublished = {\url{https://programmablesearchengine.google.com/about/}},
  note         = {Accessed September 2025},
}

@misc{chroma2025,
  author       = {Chroma Contributors},
  title        = {Chroma: Open-Source Search and Retrieval for AI},
  year         = {2025},
  howpublished = {\url{https://www.trychroma.com/}},
  note         = {Accessed September 2025},
}

@misc{serpapi2025,
  author       = {SerpApi, LLC},
  title        = {SerpApi: Real-Time Search Engine Results API},
  year         = {2025},
  howpublished = {\url{https://serpapi.com/}},
  note         = {Accessed September 2025},
}

@misc{exactmatch2025,
  author       = {Hugging Face},
  title        = {Exact Match Score: Evaluation Metric for Text Generation},
  year         = {2025},
  howpublished = {\url{https://huggingface.co/spaces/evaluate-metric/exact_match}},
  note         = {Accessed September 2025},
}

@misc{cuda_green_contexts,
  author       = {{NVIDIA Corporation}},
  title        = {{CUDA Driver API: Green Contexts}},
  howpublished = {\url{https://docs.nvidia.com/cuda/cuda-driver-api/group__CUDA__GREEN__CONTEXTS.html}},
  year         = {2025},
  note         = {Accessed: 2025-12-31}
}
}

\end{document}